\documentclass[useAMS,usenatbib]{mn2e}
\usepackage[dvips]{graphicx}
\usepackage{amssymb}
\usepackage{longtable}
\newcommand{\dn}{\hbox{$\rm D4000$}}
\newcommand{\ha}{\hbox{H$\alpha$}}
\newcommand{\hb}{\hbox{H$\beta$}}
\newcommand{\hdg}{\hbox{H$\delta_A$+H$\gamma_A$}}
\newcommand{\mgfep}{\hbox{$\rm [MgFe]^\prime$}}

\newcommand{\mgtwofe}{\hbox{$\rm [Mg_2Fe]$}}

\def\aj{AJ}%
\def\apj{ApJ}%
\def\apjl{ApJ}%
\def\apjs{ApJS}%
\def\apss{Ap\&SS}%
\def\aap{A\&A}%
\def\aaps{A\&AS}%
\def\mnras{MNRAS}%
\def\pasp{PASP}%
\title[The Ages and Metallicities of Galaxies in the Local Universe]
{The Ages and Metallicities of Galaxies in the Local Universe}
\author[A. Gallazzi et al.]{Anna Gallazzi$^{1}$\thanks{E-mail:
gallazzi@MPA-Garching.MPG.DE}, St{\'e}phane Charlot$^{1}$$^,$$^{2}$, Jarle Brinchmann$^{3}$, Simon D.M. White$^{1}$, 
\newauthor{Christy A. Tremonti$^{4}$}\\
$^{1}$Max-Planck-Institut f\"ur Astrophysik, Karl-Schwarzschild-Str. 1, 
D-85748 Garching bei M\"unchen, Germany\\
$^{2}$Institut d'Astrophysique de Paris, UMR 7095, 98 bis Boulevard Arago,
F-75014 Paris, France\\
$^{3}$Centro de Astrof{\'\i}sica da
Universidade do Porto, Rua das Estrelas - 4150-762 Porto, Portugal\\
$^{4}$Steward Observatory, University of Arizona, 933 North Cherry Avenue, Tucson, AZ 85721\\}
\begin{document}

\date{Accepted . Received ; in original form...}

\pagerange{\pageref{firstpage}--\pageref{lastpage}} \pubyear{2004}

\maketitle

  \label{firstpage}

\begin{abstract}
We derive stellar metallicities, light-weighted ages and stellar masses for a 
magnitude-limited sample of 175,128 galaxies drawn from the Sloan Digital Sky Survey Data
Release Two (SDSS DR2). We compute median-likelihood estimates of these parameters using
a large library of model spectra at medium-high resolution, covering a comprehensive range
of star formation histories. The constraints we derive are set by the simultaneous fit of 
five spectral absorption features, which are well reproduced by our population synthesis
models. By design, these constraints depend only weakly on the $\alpha$/Fe element 
abundance ratio. Our sample includes galaxies of all types spanning the full range in 
star formation activity, from dormant early-type to actively star-forming galaxies. 
By analysing a subsample of 44,254 high-quality spectra, we 
show that, in the mean, galaxies follow a sequence of increasing stellar metallicity, age
and stellar mass at increasing 4000\,\AA-break strength. For galaxies of intermediate mass,
stronger Balmer absorption at fixed 4000\,\AA-break strength is associated with higher 
metallicity and younger age. We investigate how stellar metallicity and age depend on 
total galaxy stellar mass. Low-mass galaxies are typically young and metal-poor, massive 
galaxies old and metal-rich, with a rapid transition between these regimes over the 
stellar mass range $3\times10^9 \la M_\ast \la 3\times10^{10} M_\odot$. Both high- and 
low-concentration galaxies follow these relations, but there is a large dispersion in 
stellar metallicity at fixed stellar mass, especially for low-concentration galaxies of 
intermediate mass. Despite the large scatter, the relation between stellar metallicity 
and stellar mass is similar to the correlation between gas-phase oxygen abundance and 
stellar mass for star-forming galaxies. This is confirmed by the good correlation between
stellar metallicity and gas-phase oxygen abundance for galaxies with both measures. 
The substantial range in stellar metallicity at fixed gas-phase oxygen abundance suggests
that gas ejection and/or accretion are important factors in galactic chemical evolution.
\end{abstract}


\section{Introduction}\label{intro}
The ages and metallicities of stellar populations in nearby galaxies are direct tracers
of the star formation and chemical enrichment histories of the Universe. Clues to the
ages and metallicities of the stars may be inferred from the integrated spectra of 
galaxies, using `stellar population synthesis models' \citep{tinsley78,
bruzual83,bc93, bressan94, frv97, maraston98,vazdekis99}. Analyses based on these models
have been traditionally limited by the difficulty of deriving independent constraints on
the age, star formation history, metallicity and dust content of a galaxy. For example, 
changes in age, metallicity and attenuation by dust all have similar effects on the 
colours and low-resolution spectra of galaxies. This gives rise to well-known 
`near-degeneracies' in the constraints derived on these different parameters.

The expectation is that these degeneracies may be broken, at least in part, by appealing 
to refined spectral diagnostics which are not sensitive to attenuation by dust and have 
different sensitivities to age and metallicity. Studies in this area have focused on a 
set of 25 absorption features defined and calibrated in the spectra of 460 nearby Galactic
stars obtained at Lick Observatory \citep[e.g.][]{faber73,wfg94,worthey97}. These
studies all pertain to early-type galaxies, because the lack of hot stars in the Lick
library does not allow the spectral interpretation of star-forming galaxies. The
conclusion from these studies is that, for early-type galaxies, comparisons of the 
strengths of metallic lines and age-sensitive Balmer lines {\em can} break the 
age-metallicity degeneracy, but only in a relative way: the derived ages and 
metallicities appear to depend strongly on the specific choice of metal indices 
\citep[e.g.][]{kunt01,eisenstein03, thomas04}. This is because the Galactic stars used to 
calibrate the Lick indices have approximately solar metal abundance ratios at any metallicity, 
whereas the ratio of $\alpha$-elements to iron is seen to increase from dwarf to 
massive early-type galaxies \citep[e.g.][]{wfg92}. 

A main weakness of the original calibration of Lick indices is that it relies on spectra
which were not calibrated in flux, and for which the resolution ($\sim9$\,\AA\ FWHM) is
three times lower than achieved by modern spectroscopic galaxy surveys, such as the
Sloan Digital Sky Survey (SDSS; \citealt{York}). Thus, the high-quality spectra from 
these surveys must be degraded to the calibration and resolution of the original 
Lick spectra for index-strength analyses to be performed. The situation has changed 
recently with the development of medium-high resolution ($\la3$\,\AA\ FWHM), 
flux-calibrated population synthesis models including stars in the full temperature range 
\citep{vazdekis01a, bc03}. These models can be compared directly to high-quality observed
spectra of both early-type and late-type galaxies. Several studies have also quantified 
the sensitivity of stellar absorption features to changes in element abundance ratios in
galaxy spectra \citep{tantalo98, trager00b, Thomas03a}. These developments coincide with
the advent of large homogeneous samples of galaxy spectra gathered by modern surveys such
as the SDSS.

This is the first paper of a series in which we combine modern population synthesis 
techniques with the statistical power of the SDSS to investigate the connection between
metallicity, age and stellar mass in nearby galaxies. Here, we use the medium-high 
resolution population synthesis code of \citet[ hereafter BC03]{bc03} to derive estimates of the 
metallicities, ages and stellar masses of a sample of $\sim2\times10^5$ nearby galaxies
from the SDSS Data Release Two (DR2). We adopt a Bayesian statistical approach and derive
full likelihood distributions for these physical parameters by comparing the observed
spectrum of each galaxy with a comprehensive library of model spectra corresponding to
different star formation histories. The comparison is driven by the strengths of 5 
spectral features selected to depend only weakly on the $\alpha$/Fe ratio, which we 
measure in the same way in model and observed spectra. 

An important specificity of our work is that we derive the above constraints not only for 
quiescent, early-type galaxies, but also for late-type, star-forming galaxies, for which
the contamination of stellar absorption features by nebular emission must be removed. 
We explore the relationships between metallicity, age and stellar mass and the dependence
of these relationships on galaxy structure. In a companion paper, we exploit these results
to investigate the physical origin of the colour-magnitude relation and of the relation
between $\rm Mg_2$-index strength and velocity dispersion for early-type galaxies. The
total metal content of the local Universe and the distribution of metals as a function of
galaxy properties will be the subject of a subsequent paper.

The paper is organized as follows. In Section \ref{approach} below, we present our sample
of SDSS spectra and the models used to interpret them. We also describe our method for 
deriving metallicities, ages and stellar masses from observed galaxy spectra. The results
are presented in Section \ref{results}, where we study the dependence of metallicity and 
age on total stellar mass and the age-metallicity relation as a function of galaxy 
structure. We also address in that section the influence of aperture bias on our derived
parameters. We summarize and briefly discuss our results in Section \ref{summary}.

\section{The approach}\label{approach}
In this section, we describe our approach for deriving estimates of light-weighted ages
and metallicities from the observed spectra of SDSS galaxies. These spectra are discussed
in Section \ref{spectra}, along with the models we use to interpret them. In Section 
\ref{measure}, we select a set of spectral absorption features which we argue should most
robustly constrain age and metallicity. Then, in Section \ref{method}, we outline the 
statistical approach we adopt to derive ages and metallicities from galaxy spectra. We
also highlight the sensitivity of our constraints on age and metallicity to the 
observational signal-to-noise ratio (S/N) of the spectra.

\subsection{Observed and model spectra}\label{spectra}
The observed spectra we consider are drawn from the SDSS DR2 \citep{dr204}. The SDSS
is an imaging and spectroscopic survey of the high Galactic latitude sky, which will
obtain $u$, $g$, $r$, $i$ and $z$ photometry of almost a quarter of the sky and spectra
of at least 700,000 objects \citep{York}. The spectra are taken using 3\arcsec-diameter
fibres, positioned as close as possible to the centres of the target galaxies. The flux-
and wavelength-calibrated spectra cover the range from 3800 to 9200\,\AA\ with 
a resolution of $\sim 1800$. Our sample, drawn from the SDSS DR2, includes unique spectra 
of 196,673 galaxies with Petrosian $r$-band magnitudes in the range $14.5<r<17.77$ (after correction 
for foreground Galactic extinction using the extinction maps of \citealt{Schlegel}). The 
galaxies span the full range of types, from actively star-forming, late-type galaxies to 
dormant, early-type galaxies. The median redshift is 0.13. A more detailed description
of this sample is given in Section \ref{results} below.

To interpret these observed spectra in terms of physical parameters such as age, 
metallicity and stellar mass, we use the recent population synthesis models of BC03. 
These models are based on `STELIB', a newly available library of observed stellar spectra 
assembled by \cite{leborgne}. The models have a spectral resolution of 3\,\AA\ FWHM 
across the whole wavelength range from 3200 to 9500\,\AA. They are thus ideally matched
to the SDSS spectra. 

We note that, to interpret the spectra of star-forming and active galaxies with these 
models, we must first remove the contamination of the observed spectra by nebular 
emission lines. This is achieved using the procedure outlined by \cite{christy04}, which
is optimized for use with SDSS galaxy spectra \citep[see also][]{christyPhD}. This consists 
in performing first a non-negative least-squares fit of the emission-line-free regions of the 
observed spectrum, using a set of model template spectra broadened to the observed velocity
dispersion (the template spectra correspond to 30 instantaneous-burst models of different
ages and metallicities computed using the BC03 code). Once the fitted spectrum is 
subtracted from the observed spectrum, the residuals can be fitted to Gaussian-broadened
emission-line templates. The method assumes a single broadening width for all the Balmer
lines, and another (independent) width for all the forbidden lines.  The strength of 
each line is fitted independently. Then, the fitted emission lines are subtracted from the
original observed spectrum to produce a `pure' absorption-line spectrum suited to our 
analysis.

\subsection{Stellar absorption diagnostics of age and metallicity}\label{measure}
The strongest stellar absorption features in the optical spectra of galaxies form
the basis of the Lick system of spectral indices (\citealt{burstein84,gorgas93,wfg94,
worthey97,trager98}; see also Section~\ref{intro}). Each index in this system is defined
by a central `feature bandpass' and two adjacent `pseudo-continuum bandpasses'. The above
studies have shown that some Lick indices are primarily sensitive to age, such as those
based on H-Balmer lines, while others are primarily sensitive to metallicity, such as
several Fe- and Mg-based indices at wavelengths between 4500 and 5700\,\AA. We also
consider here the 4000-\AA\ break index of \citet{balogh99}, which we denote \dn, which is 
defined as the ratio of the average flux densities in the narrow bands 4000--4100~{\AA}
and 3850--3950~{\AA}.\footnote{This index is sometimes denoted by $\rm D_n(4000)$.}
This index depends somewhat on metallicity but correlates more with the ratio of present
to past-averaged star formation rates in galaxies (see fig.~2 of \citealt{kauf03a} and 
fig.~27 of \citealt{jarle03}).

In previous studies based on models with low spectral resolution, the strengths of Lick
indices had to be modelled analytically as functions of the effective temperatures, 
gravities and metallicities of the stars (\citealt{wfg94,worthey97,gorgas99}). These `fitting
functions' were not appropriate for hot stars, and hence, index-strength analyses
had to be restricted to old stellar populations. Moreover, as mentioned in 
Section~\ref{intro}, the stellar spectra on which the Lick indices were originally 
calibrated had lower resolution than typical galaxy spectra today and were not flux-calibrated. 
These various weaknesses are resolved here by our adoption of the medium-high resolution, 
flux-calibrated BC03 models. These models can be compared directly to the SDSS spectra of 
galaxies with any star formation history, and the Lick indices can be measured in the same way in model 
and observed spectra. 

Like previous models, however, the BC03 models rely on a spectral library of nearby
stars with near solar metal abundance ratios at any metallicity. The models are therefore
expected to show discrepancies when compared to galaxies where the abundance ratios 
differ from those of nearby stars.\footnote{\label{alphafe}The abundance ratio of 
$\alpha$ elements (such as N, O, Mg, Ca, Na, Ne, S, Si, Ti), which are produced mainly 
by Type II supernovae, to Fe-peak elements (such as Cr, Mn, Fe, Co, Ni, Cu, Zn), which
are produced mainly by Type Ia supernovae, is observed to vary in external galaxies 
\citep[e.g.,][]{wfg92,thomas03c}.} In fact, such discrepancies appear to be responsible
for the fact that some spectral features, such as CN$_1$, CN$_2$, TiO$_1$, TiO$_2$, Ca4227
and several Mg and Fe lines, were not well reproduced when BC03 compared their models 
with a sample of high-quality galaxy spectra drawn from SDSS Early Data Release 
\citep{stoughton}.

Several studies have addressed the dependence of Lick index strengths on changes in the 
relative ratios of heavy elements \citep[][ see also BC03]{gonzalez93a,tripicco95,
tantalo98, trager00b,vazdekis01a,Thomas03a,tantalo04,thomas04}. These studies have led to the 
identification of composite Mg+Fe indices, which are sensitive to metallicity (i.e. the fraction 
by mass of all elements heavier than helium over the total gas mass) but show little
sensitivity to $\alpha$/Fe (i.e. the ratio of the total mass of $\alpha$ elements to the
mass of iron; see footnote~\ref{alphafe}). Among these, we use here 
\begin{equation}
\mgfep = \rm \sqrt{Mgb~(0.72~Fe5270+0.28~Fe5335)},
\end{equation}
as proposed by \citealt{Thomas03a}, and
\begin{equation}
\mgtwofe = \rm 0.6~Mg_2 + 0.4~log(Fe4531+Fe5015), 
\end{equation}
as defined in BC03.
 
We wish to extract ages and metallicities from the SDSS galaxy spectra. We therefore 
fit simultaneously both metal-sensitive and age-sensitive indices. Among the 28 spectral 
indices studied by BC03, we concentrate on those that are best reproduced by the models. 
This requirement already excludes several Fe-based indices, the three Mg-based indices, 
Ca4227, Ca4455, the CN, TiO and NaD features (see fig.~18 of BC03). 
Among the remaining indices we identify those that are known to have at most a weak dependence 
on $\alpha$/Fe. \mgtwofe\ and \mgfep\ are suitable metal-sensitive indices 
and \hb\ is a suitable age-sensitive index. We also include \dn, which is sensitive to the ratio 
of present to past-averaged star formation rate (see above). We are not aware of any study 
indicating the dependence of this index on $\alpha$/Fe. Finally, to better constrain age we decided to 
include also the two higher-order Balmer lines, H$\delta_A$ and H$\gamma_A$. These indices have 
been recently shown to depend on variations in element abundance ratios at metallicities around 
solar and above \citep{thomas04,korn05}. However, we find that including them in our procedure 
does not produce results systematically different from those obtained without them, while 
it provides smaller errors on both age and metallicity estimates. Therefore, we decide 
to use also \hdg. We choose the sum of H$\delta_A$ and H$\gamma_A$ because it is better 
reproduced by the models than the two indices separately (see fig.~18 of BC03). 
Thus, our final set of indices is composed of \dn, \hb, \hdg, \mgtwofe\ and \mgfep. 
This is the minimum set that allows us to derive good constraints on 
metallicity and age simultaneously and to recover well the parameters of simulated galaxies  
(see section~\ref{accuracy} below).

%To extract ages and metallicities from the SDSS galaxy spectra, we select a set of 
%indices which are sensitive to changes in age and metallicity, but have at most a weak  
%dependence on $\alpha$/Fe. Specifically, we adopt \hb\ and \hdg\ as age-sensitive indices
%(we choose the sum of H$\delta_A$ and H$\gamma_A$ because it is better reproduced by
%the models than the two indices separately; see fig.~18 of BC03). We
%also include \dn, which is sensitive to the ratio of present to past-averaged star 
%formation rate (see above). Finally, we adopt \mgtwofe\ and \mgfep\ as 
%metal-sensitive indices.

\begin{figure*}
\centerline{\includegraphics[width=13truecm]{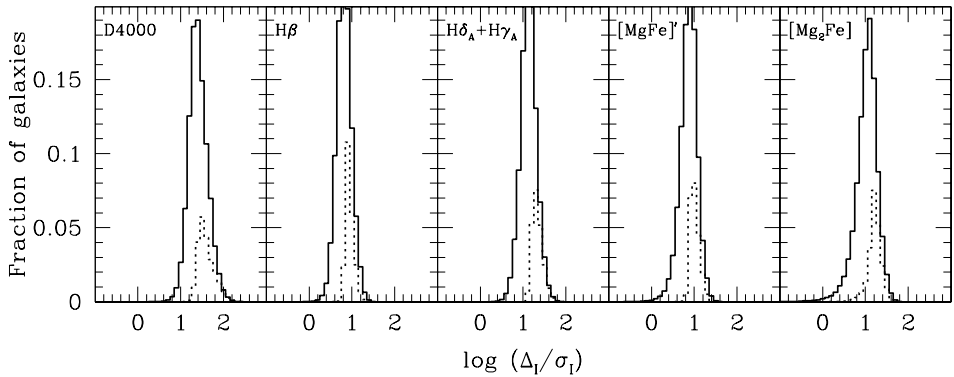}}
\caption{Distributions in `resolving power' of the 5 spectral features selected to
constrain the ages and metallicities of SDSS galaxies, as indicated. The resolving power
is defined as the ratio between the 5\%--95\% percentile range $\Delta_I$ of index 
strengths in the sample and the observational error $\sigma_I$ of each galaxy. In each 
panel, the solid histogram shows the distribution for our sample of 196,673 SDSS-DR2 galaxies, 
and the dotted histogram the distribution for the subsample of 44,347 galaxies with mean 
signal-to-noise per pixel greater than 20.}\label{fig1}
\end{figure*}

The solid histograms in Fig.~\ref{fig1} show the distributions in `resolving power'
of these five spectral features for the galaxies in our sample. The resolving power is 
defined as the ratio between the 5--95 percent percentile range of the distribution 
of index strengths for all galaxies in the sample, $\Delta_I$, and the observational error 
for each galaxy, $\sigma_I$. The resolving power is largest ($\sim30$) for \dn\ and
slightly lower ($\sim10$) for the other indices. Also shown as dotted   
histograms in Fig.~\ref{fig1} are the analogous distributions in resolving power for 
a subsample of galaxies with mean signal-to-noise per pixel greater than 20. As expected, 
the distributions for these galaxies are shifted to higher values, because of the smaller 
observational errors.

We note that the strengths of some spectral absorption indices are sensitive to the
stellar velocity dispersion in a galaxy \citep[e.g.][]{davies93,longhetti98,trager98,kunt04}. 
The indices that are most affected are those measured with the narrowest pseudo-continuum 
bandpass definitions. In particular this effect is seen in Fe-based indices and therefore 
also in the composite Mg+Fe indices. The BC03 population synthesis code provides SSP spectra 
broadened to different velocity dispersions. This allows us to compare each galaxy 
spectra with models that have a similar velocity dispersion (see Section~\ref{method}).

\subsection{Statistical estimates of age and metallicity}\label{method}
We wish to estimate not only the most likely values of the ages and metallicities of SDSS
galaxies, but also the accuracy of these values. To this purpose, we adopt a Bayesian 
statistical approach, similar to the one outlined by \cite{kauf03a}. The goal is to obtain
the likelihood distribution of a given parameter $X$ in the space of all possible values
of $X$. This is obtained by comparing the observational data with a set of models that 
populate the space of all possible $X$ according to a prior distribution which represents
our prejudice about the relative likelihood of different $X$ values in absence of any data.

We generate a library of 150,000 Monte Carlo realizations of a full range of physically
plausible star formation histories. Following \cite{kauf03a}, each star formation history
is parametrized in terms of a continuous model in which stars are formed from the time
$t_{\rm form}$ to the present according to the law $\psi(t) \propto \exp(-\gamma t)$. 
Random bursts of star formation are superposed on this continuous model. We take the 
formation time $t_{\rm form}$ to be uniformly distributed between 13.5 and 1.5 Gyr and the 
star formation time-scale parameter $\gamma$ to be uniform over the interval from 0 to 1 
Gyr$^{-1}$. The bursts can occur at all times after $t_{\rm form}$ with equal probability,
set in such a way that 10 percent of the galaxies in the library experience a burst in the
last 2 Gyr.\footnote{In \cite{kauf03a} this parameter was set such that 50 percent of the models 
in the library experienced a burst in the last 2 Gyr. We reduce this fraction to 10 percent because 
it provides a more uniform distribution of the models in light-weighted age. The influence 
on our results from changing this fraction is discussed in Section~\ref{accuracy2}.} They are 
parametrized in terms of the fraction $A$ of stellar mass produced 
during the burst relative to the total mass formed by the continuous model. The ratio $A$
is logarithmically distributed between 0.03 and 4. During a burst, stars form at a 
constant rate for a time distributed uniformly in the range $3\times10^7$--$3\times10^8$
yr. The velocity dispersions of the models are distributed uniformly in the range
50--350 $\rm km~s^{-1}$. We further take the models to be distributed logarithmically in 
metallicity in the range 0.2--2.5\,$Z_\odot$ and make the density of models drop smoothly
as $(\log Z)^{1/3}$ at metallicities from 0.2 down to 0.02\,$Z_\odot$, in order not to 
overrepresent extremely metal-poor models. All stars in a given model have the same
fixed metallicity, which we interpret as the `(optical) light-weighted' metallicity. 

For each model in the library we compute the following properties:
\begin{enumerate}
\item
the strengths of the \dn, \hb, \hdg, \mgtwofe\ and \mgfep\ spectral indices,
measured in the same way as in the SDSS spectra;

\item
the $r$-band light-weighted age, evaluated by the integral 
$t_r=\int_0^t \left[d\tau\,\psi(t-\tau)\,f_r(\tau)\,\tau \right]
    /\int_0^t \left[d\tau\,\psi(t-\tau)\,f_r(\tau) \right]$, 
where $f_r(\tau)$ is the total $r$-band flux produced by stars of age $\tau$. We
refer below to the $r$-band light-weighted age simply as the `age' of a galaxy;
\item
the $z$-band stellar mass-to-light ratio $M_\ast/L_z$, which accounts for the gas
mass returned to the ISM by evolved stars (see also section 3.1 of BC03);
\item
the apparent $u$, $g$, $r$, $i$ and $z$ magnitudes of the model at redshifts between
0 and 0.3 in steps of 0.01. 
\end{enumerate}

\begin{figure*}
\centerline{\includegraphics[width=13truecm]{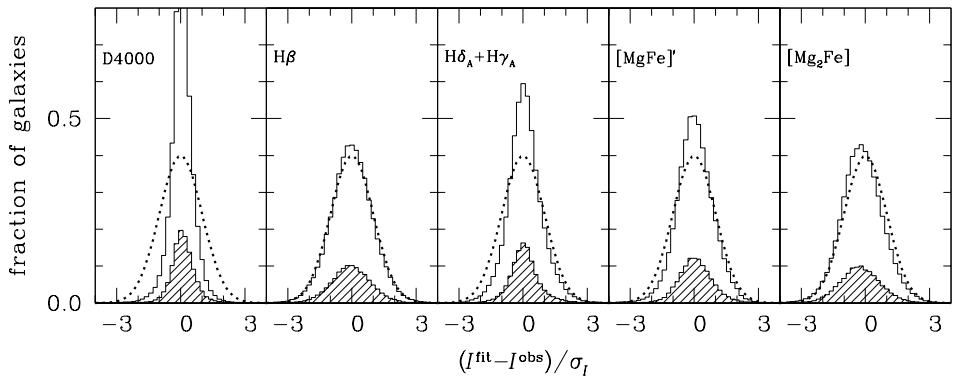}}
\caption{Simultaneous fit of the 5 spectral features chosen to derive age and metallicity
estimates from SDSS galaxy spectra using the model library described in 
Section~\ref{method}. The histograms show the distributions of the differences between 
the best-fit ($I^{\rm fit}$) and observed ($I^{\rm obs}$) strengths of each index (as 
indicated), in units of the observational error ($\sigma_I$). The plain histograms 
are for the full sample of 196,673 galaxies, while the hatched histograms are for the 
subsample of 44,347 galaxies with mean S/N per pixel greater than 20. Only those models
for which the stellar velocity dispersion is within $\rm \pm 15~km~s^{-1}$ of the 
observed one are included in the fit for each galaxy. For reference, the dotted line in
each panel shows a Gaussian distribution with unit standard deviation.}
\label{fig2}
\end{figure*}

The models in the library provide accurate simultaneous fits to the strengths of the 5
spectral indices \dn, \hb, \hdg, \mgtwofe\ and \mgfep\ that we have selected
to derive age and metallicity estimates from SDSS galaxy spectra. To illustrate this, 
Fig.~\ref{fig2} shows the distribution of the differences between the best-fit and 
observed strengths of each index, in units of the observational error, for the galaxies
in our sample. Because of the dependence of the composite Mg+Fe indices on stellar 
velocity dispersion, when fitting the index strengths of an observed galaxy, we only 
include those models for which the stellar velocity dispersion is within $\rm \pm 
15~km\,s^{-1}$ of the observed one.\footnote{The median uncertainty on velocity 
dispersion is $\rm \sim15~km\,s^{-1}$ for the full sample and only $\rm \sim7~km\,s^{-1}$ 
for galaxies with S/N$>$20.} The plain histograms in Fig.~\ref{fig2} show
the distributions for the sample as a whole, while the hatched histograms show the 
distributions for those galaxies with a median S/N per pixel greater than 20. For
both samples, the deviations between observations and best-fit models are within the 
observational errors for all indices, as indicated by the comparison with a Gaussian 
distribution of unit standard deviation (dotted line). There is no strong correlation 
between the residuals of the different indices.

The comparison of the strengths of \dn, \hb, \hdg, \mgtwofe\ and \mgfep\ in
the spectrum of an observed SDSS galaxy with the strengths of these indices in every 
model spectrum in the library allows us to construct the probability density functions 
(PDFs) of physical parameters (such as age, metallicity and mass-to-light ratio) for
that galaxy. This is achieved by assigning to each model a weight $w=\rm exp(-\chi^2/2)$,
where $\chi^2$ is calculated by comparing the strengths of the 5 indices measured in the
observed spectrum with those measured in the model spectrum, given the observational
measurement errors. The PDF of a selected physical parameter is then simply given by the
distribution in that parameter of the weights $w$ of all the models in the library. 

The PDF of a given parameter can then be characterized by its mode, which corresponds to
the most likely value of the parameter, its median, which can differ from the mode for 
non-symmetric distributions, and by a confidence interval within which the parameter is
constrained at a certain probability level. We often quote below the 68 percent 
confidence interval corresponding to the 16\%--84\%  percentile range of the PDF, which
would be equivalent to the $\pm 1\sigma$ range for a Gaussian distribution.

\subsection{Accuracy of the estimates}\label{accuracy}
\subsubsection{Dependence on observational properties}\label{accuracy1}
We now want to illustrate the kind of constraints that can be obtained on the ages and
metallicities of SDSS galaxies with different spectral properties using the method 
outlined above. \citet{kauf03a} have shown that the combination of H-Balmer lines and \dn\
is a discriminating diagnostic of the recent star formation activity in galaxies. We
therefore draw 4 galaxies from our sample with high-quality spectra (median S/N per
pixel larger than 30) located at different positions along the sequence occupied by SDSS
galaxies in the \hdg\ versus \dn\ diagnostic diagram. 

\begin{figure*}
\centerline{\includegraphics[width=8truecm]{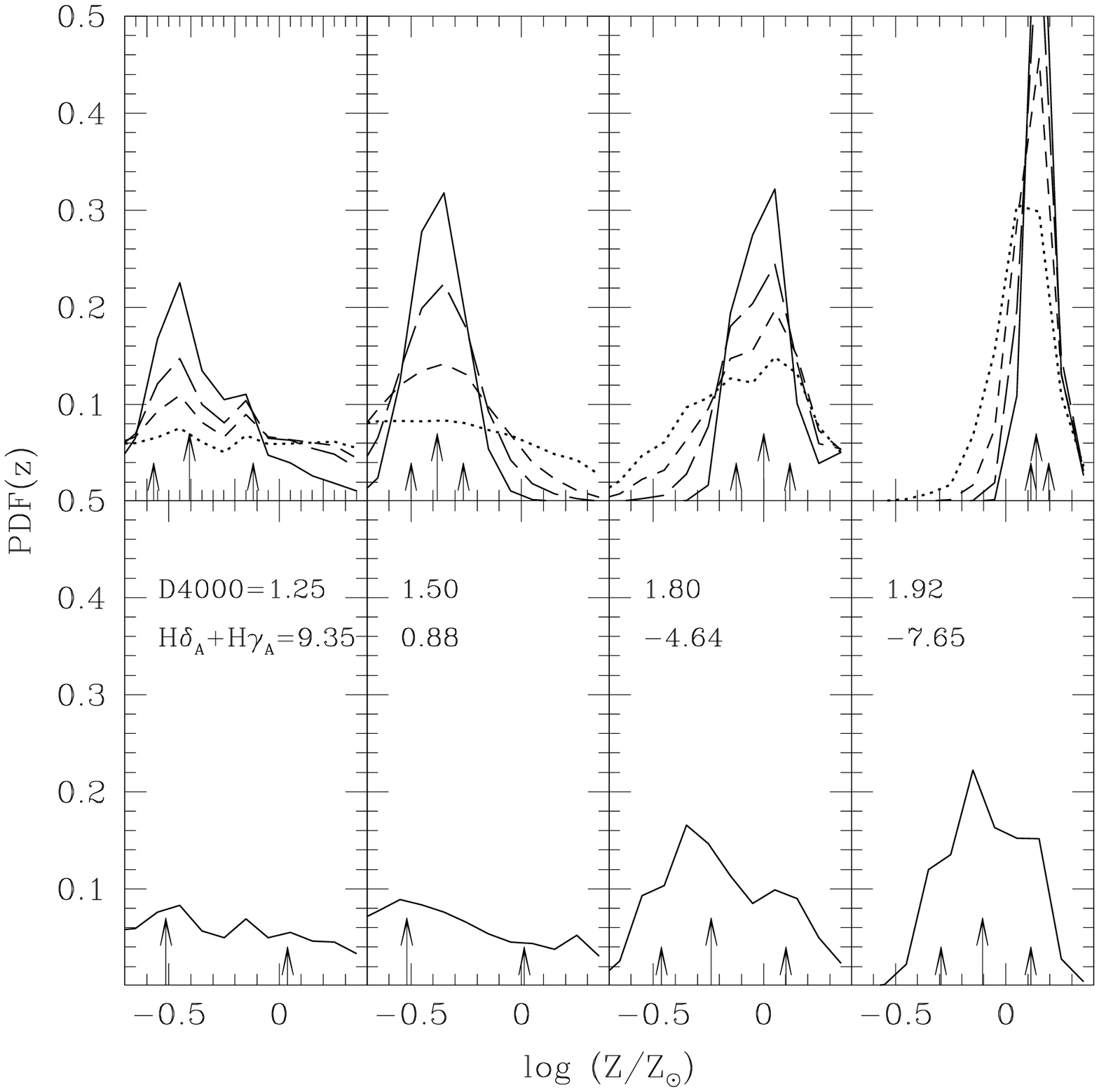}
\includegraphics[width=8truecm]{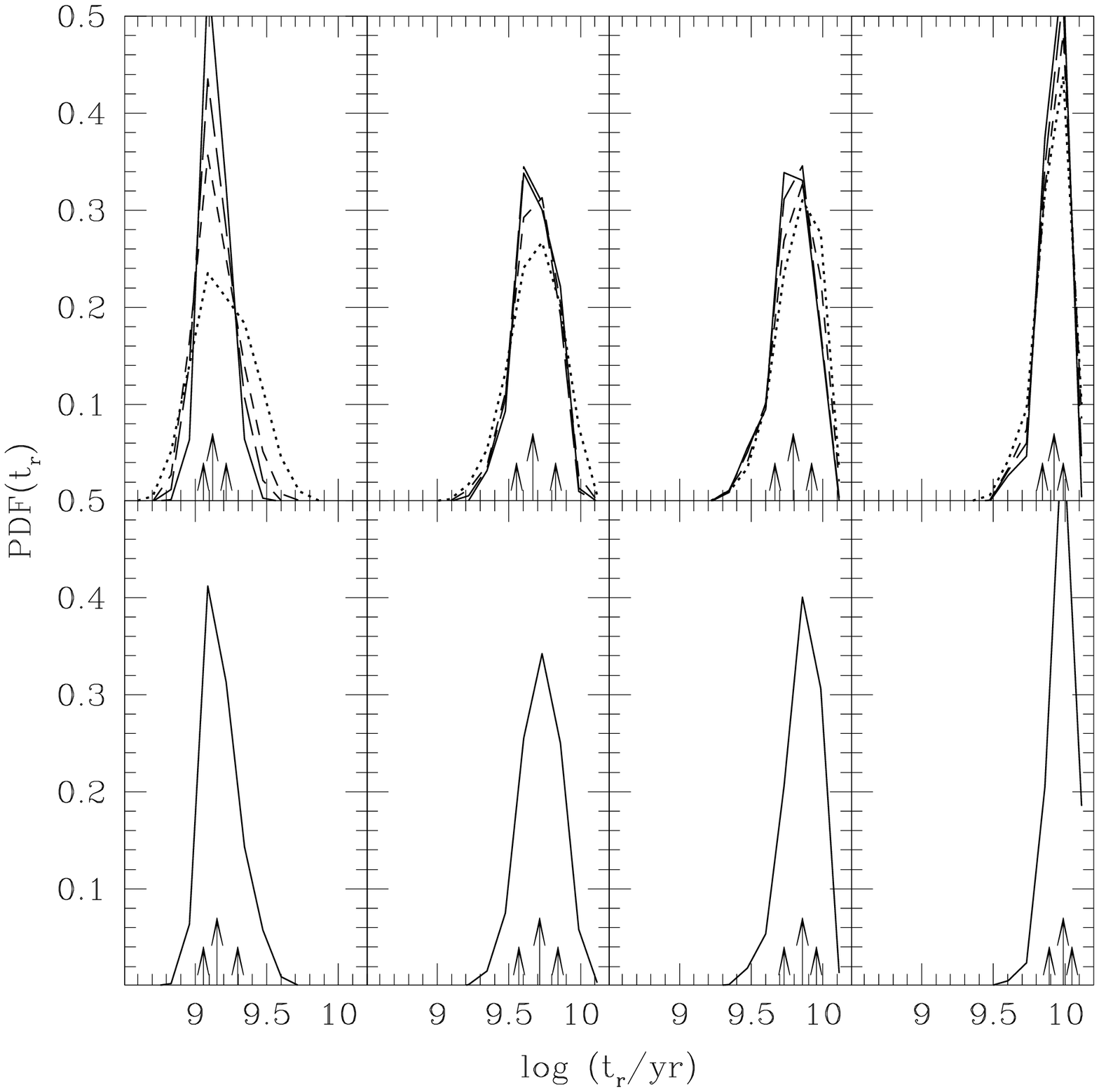}}
\caption{Probability density functions of stellar metallicity (left-hand plot) and 
$r$-band light-weighted age (right-hand plot) for 4 SDSS galaxies with high-quality
spectra (median S/N per pixel larger than 30) and different \dn\ and \hdg\ strengths
(indicated in the bottom panels of the left-hand plot). The solid PDFs in the bottom 
panels were obtained when including only the age-sensitive indices \dn, \hb\ and \hdg\
to constrain the fits. Those in the top panels were obtained after including also the 
metal-sensitive indices \mgtwofe\ and \mgfep. In each panel, the arrows indicate
the median (longer one) and the $\rm 16^{th}$ and $\rm 84^{th}$ percentiles (shorter ones) of the PDF.
The long-dashed, short-dashed and dotted PDFs in the top panels show the constraints 
obtained when degrading the original galaxy spectra to a median S/N per pixel of 30, 20
and 10, respectively (see text for detail).}
\label{fig3} 
\end{figure*}

The solid distributions in Fig.~\ref{fig3} show the constraints obtained on the 
metallicities (left-hand plot) and ages (right-hand plot) of these galaxies in two cases:
when including only the age-sensitive indices \dn, \hb\ and \hdg\ to constrain the fits
(bottom panels), and when including also the metal-sensitive indices \mgtwofe\
and \mgfep\ (top panels). As expected, the ages are well constrained by \dn, \hb\ and 
\hdg\ alone, and the corresponding PDFs do not change appreciably when including also 
the constraints from metal-sensitive indices. In contrast, the metallicities are well 
constrained only when the metal-sensitive indices are included in the fit. In 
each panel in Fig.~\ref{fig3}, the arrows indicate the median (longer one) and the $\rm
16^{th}$ and $\rm 84^{th}$ percentiles (shorter ones) of the PDF. The 68 percent confidence 
interval becomes narrower when all the five indices are included, and the median of the 
distribution in this case is consistent with that obtained when including only the 
age-sensitive indices. We note that age and metallicity appear to correlate with \dn\ 
(indicated on the figure) for these 4 galaxies, and that the constraints on metallicity 
are weakest for the galaxy with the lowest \dn. This is not a coincidence, as we shall 
see in Section~\ref{results} below.

It is important to note that the observed S/N directly influences the uncertainties in 
the age and metallicity estimates. To investigate this, we consider a set of galaxies 
with index strengths similar to those of the galaxies in Fig.~\ref{fig3}, but with
lower median S/N per pixel, e.g. 20. We take the average errors in the index strengths 
of these low-S/N galaxies to be typical of errors that would be measured at the same S/N
for the galaxies in Fig.~\ref{fig3}. We then mimic 100 realizations of `degraded
spectra' of the high-S/N galaxies by randomly drawing index strengths from Gaussian 
distributions centered on their index values, of widths the typical errors obtained from
the lower-S/N galaxies. The PDFs of age and metallicity may be computed for each 
realization, and the average PDF of the 100 realizations is a good estimator of the 
results that would be obtained at low S/N for these galaxies. 

In the upper panels of Fig.~\ref{fig3}, we show the average PDFs obtained in this
way when the median S/N per pixel of each galaxy is degraded to 30 (long-dashed line), 20
(short-dashed line) and 10 (dotted line). A median S/N per pixel of 10 is not sufficient
to constrain metallicity well (except for the most metal-rich galaxy, which 
has the spectrum with the strongest absorption features). As we increase the
S/N, the distributions narrow down and converge to the PDFs obtained from the original
high-S/N spectra. We conclude that a median S/N per pixel of at least 20 is required to
reliably constrain metallicity. This is higher than the median value of $\sim15$ for the
SDSS-DR2 sample. The age estimates do not appear to be significantly affected by low S/N
(except for the youngest galaxy), probably because they are constrained strongly by \dn,
which has small errors. The results of this exercise are summarized in 
Table~\ref{error_stat}.

\begin{table}
\caption{Median-likelihood estimates of metallicity and age for the four galaxies shown in 
Fig.~\ref{fig3}, obtained by degrading the actual signal-to-noise ratio of the spectrum 
($\rm S/N_{obs}$) to 30, 20 and 10. The quoted errors are one half the 68 percent confidence 
range of the PDFs in the upper panels of Fig.~\ref{fig3}. Each column corresponds to a galaxy, 
ordered from left to right as in Fig.~\ref{fig3}.}\label{error_stat}
\centering
\begin{tabular}{|l|c|c|c|c|}
\multicolumn{5}{|c|}{}\\
\hline
% & \multicolumn{1}{c}{$\dn=1.25$} & \multicolumn{1}{c}{$\dn=1.50$} &\multicolumn{1}{c}{$\dn=1.80$}
% &\multicolumn{1}{c}{$\dn=1.92$} \\
%\dn & 1.25 & 1.50 & 1.80 & 1.92 \\
%\hline
\multicolumn{5}{|c|}{$\log (Z/Z_\odot)$} \\ 
\hline
$\rm S/N=10$ 	& $-$0.43$\pm$0.57 & $-$0.42$\pm$0.43 & $-$0.07$\pm$0.27 &  0.09$\pm$0.12 \\
$\rm S/N=20$ 	& $-$0.36$\pm$0.44 & $-$0.39$\pm$0.27 & $-$0.02$\pm$0.21 &  0.12$\pm$0.08 \\
$\rm S/N=30$ 	& $-$0.34$\pm$0.35 & $-$0.37$\pm$0.17 &  0.00$\pm$0.16 &  0.13$\pm$0.06 \\
$\rm S/N_{obs}$ & $-$0.41$\pm$0.22 & $-$0.38$\pm$0.12 &  0.00$\pm$0.12 &  0.14$\pm$0.04 \\
\hline
\multicolumn{5}{|c|}{$\log (t_r/yr)$}\\
\hline
$\rm S/N=10$ 	& 9.19$\pm$0.21 &  9.68$\pm$0.18 &  9.84$\pm$0.15 &  9.92$\pm$0.11 \\
$\rm S/N=20$ 	& 9.12$\pm$0.15 &  9.68$\pm$0.15 &  9.83$\pm$0.14 &  9.92$\pm$0.10 \\
$\rm S/N=30$ 	& 9.12$\pm$0.12 &  9.66$\pm$0.13 &  9.79$\pm$0.14 &  9.92$\pm$0.09 \\
$\rm S/N_{obs}$ & 9.12$\pm$0.08 &  9.66$\pm$0.13 &  9.79$\pm$0.13 &  9.92$\pm$0.07 \\
\hline
\end{tabular}
\end{table}

%\begin{table}
%\caption{}\label{error_stat2}
%\centering
%\begin{tabular}{l|cc}
%\hline
% & $\langle \sigma_{\log Z}\rangle$ & $\langle \sigma_{\log t_r}\rangle$ \\
%\hline
%All 			    & 0.20 & 0.14 \\
%$\rm S/N\geq20$ 	    & 0.12 & 0.12 \\
%$\rm S/N\geq20$, $C\geq2.8$ & 0.10 & 0.11 \\
%$\rm S/N\geq20$, $C\leq2.4$ & 0.19 & 0.13 \\
%\hline
%\end{tabular}
%\end{table}

The results of Fig.~\ref{fig3} would not change dramatically if we restricted
the fit to only a subset of the selected spectral indices. 
%This is because the indices used in our
%analysis are sensitive to age and metallicity, but only weakly to the $\alpha$/Fe ratio 
%(Section~\ref{measure}). 
We have checked this by using only \hb\ to derive age estimates
for the galaxies in Fig.~\ref{fig3}, and only \mgfep\ to derive metallicity 
estimates. The median-likelihood estimates of age and metallicity obtained in this way
fall within the $1\sigma$ errors of those shown in Fig.~\ref{fig3}. The main effect
of reducing the number of indices included in the fitting is to broaden the derived 
likelihood distributions. For the galaxies of Fig.~\ref{fig3}, the average error on age 
increases from 0.10 dex to 0.14 dex when fitting \hb\ alone\footnote{It has been noticed that 
higher-order Balmer lines can yield younger ages than those predicted by \hb, presumably 
because of the dependence of higher-order lines on $\alpha$/Fe \citep[e.g.][]{kunt00,eisenstein03}. 
However, we do not see any significant difference in the results when using either \hdg\ only 
or \hb\ only.}, 
while the average error on metallicity increases from 0.13 dex to 0.29 dex when fitting 
\mgfep\ alone. 

\subsubsection{Possible systematic uncertainties}\label{accuracy2}
To identify any potential bias in our method, we also tested how well it can recover
ages and metallicities of model galaxies for which these parameters are known.
We selected 3000 models at random from the library described in Section~\ref{method}.
We added Gaussian noise to the index strengths of these models to reflect
the average observational errors of our SDSS sample. The ages and metallicities 
recovered by our method for these models showed no systematic deviation from the true 
values. For both age and metallicity, the deviations $\Delta \log Z$ and $\Delta
\log t_r$ for the 3000 models followed Gaussian distributions centred on zero of
width $\sim0.15$. Interestingly, the deviations in age and metallicity appear to 
correlate with each other with a slope $\Delta \log Z/\Delta\log t_r \approx -0.74$.
This is consistent with the age-metallicity degeneracy identified in Section~\ref{agez}
below. We have checked that the way in which $\Delta \log Z$ and $\Delta \log t_r$ 
respond to (noise-induced) changes in metal-sensitive and age-sensitive indices
for galaxies with different star formation histories cannot lead to spurious
correlations between index strengths and metallicity/age residuals.

We mentioned above that higher-order Balmer lines may be sensitive to 
the $\alpha$/Fe ratio. To quantify the potential error on our metallicity and 
age estimates for galaxies with enhanced $\alpha$/Fe relative to solar, we compared 
the predictions of the \cite{thomas04} stellar populations models at solar element abundance ratio, 
$\rm\alpha/Fe=0$, with those for $\rm\alpha/Fe=0.3$ (a typical ratio for massive elliptical galaxies). 
We chose 3000 models at random from our library and perturbed their index strengths with Gaussian noise, 
as above. In addition, we increased the values of H$\delta_A$ and H$\gamma_A$ to reflect  
the difference between the $\rm\alpha/Fe=0$ and $\rm\alpha/Fe=0.3$ tracks, interpolating 
in metallicity and age to the values for each model considered. The distributions in 
$\Delta \log Z$ and $\Delta \log t_r$ are well represented by Gaussian centred on $+0.05$ 
and $-0.05$, respectively, of width given by the average error on metallicity and age 
(0.2 and 0.13 respectively). This test indicates that we tend to overestimate the stellar 
metallicities and underestimate the light-weighted ages of galaxies with supersolar abundance 
ratios by $\sim0.05$ dex. However, this seems in contradiction with the fact that the results 
on the galaxies in our sample do not vary systematically if we include or exclude \hdg\ in the 
fit. This is true also for massive early-type galaxies, which are likely to be $\alpha$-enhanced 
\citep[e.g.][]{wfg92}. The offset of $\sim0.05$ dex may thus be regarded as an upper limit to the 
error in the ages and metallicities derived from our analysis in the case of non-solar abundance ratios. 

Another possible source of systematic error is the choice of prior according to which 
our model library populates the parameter space. In particular, the mix of continuous and 
bursty star formation histories may influence the physical parameters in which we are 
interested, mainly the light-weighted age. To test for this effect, we generated a Monte Carlo 
library with a modified prior, by increasing to 50 percent (instead of 10 percent) the fraction 
of models that can undergo a burst of star formation in the last 2 Gyrs. We then compared the 
(median-likelihood) estimates of light-weighted age and stellar metallicity 
derived with this modified prior and with our standard prior. Increasing the fraction of bursts,
we derive ages $\sim$0.07 dex younger, on average, than those derived with our standard prior. 
This bias mainly affects old, early-type galaxies. Similarly the metallicities are on average 
0.04 dex higher than those derived with the standard prior. The effect of these offsets on the 
relations between age, metallicity and stellar mass that we discuss in Section~\ref{res2} 
is very small. The zeropoint of the relations changes according to the offsets in metallicity 
and age reported above, but the shape of the relations remains identical.

\subsubsection{Stellar mass estimates}\label{accuracy3}
In the remainder of this paper, we will be interested in the dependence of age and 
metallicity not only on directly observed properties,  such as spectral features and 
morphology, but also on `derived' quantities, such as stellar mass. \cite{kauf03a} derived
Bayesian likelihood estimates of the stellar masses of a sample of SDSS galaxies, based on
fits of the $\rm H\delta_A$ and \dn\ absorption indices. We use a similar approach here 
and estimate stellar masses for all the galaxies in our sample, based on the fits of \dn,
\hb, \hdg, \mgtwofe\ and \mgfep. We compute the PDF of the stellar mass for each
galaxy by scaling the $z$-band mass-to-light ratio $M_\ast/L_z$ of each model to the 
observed, total $z$-band luminosity of the galaxy.\footnote{We assume that the $M_\ast/
L_z$ ratio is the same for the whole galaxy as it is in the region sampled by the fibre. 
See Fig.~\ref{fig13} for the typical fraction of light that enters the fibre.}
We compute $M_\ast/L_z$ at the observed galaxy redshift and include the effects of 
attenuation by dust. The $z$-band attenuation $A_z$ is inferred from the difference 
between the emission-line corrected fibre $r-i$ colour\footnote{Fibre magnitudes are 
obtained using the SDSS photometry directly out to the radius covered by the fibre. They
are thus directly comparable to quantites derived from spectroscopy, provided that the 
fibre is positioned at the centre of the galaxy. Corrections for emission lines are 
obtained by comparing the magnitudes measured off the spectrum before and after removing
emission lines.} of the galaxy and the $r-i$ colour of the redshifted (dust-free) model,
assuming a single power law ($\propto\lambda^{-0.7}$) attenuation curve \citep{cf00}. 

The true attenuation can of course not be negative, but imposing strictly $A_z > 0$
would not account for the errors affecting the measurements of indices and magnitudes.
Moreover, at high metallicities, there is a potential problem that the dust-free models 
providing the best fits to the observed absorption-line strengths can be {\em redder} than the
observed galaxy (implying negative apparent attenuations). This problem arises because of
a discrepancy between the spectral absorption features and the integrated colours of the
BC03 models at the metallicity $2.5\,Z_\odot$. The colours of these models correspond to 
stellar evolutionary tracks with (and colour-temperature calibrations for) $Z=2.5\,
Z_\odot$, but the absorption-line strengths had to be calibrated using stellar spectra 
for slightly lower metallicity, $1.6\la Z/Z_\odot\la 2$ (see Appendix A of BC03). 
Hence, the metallicity scale, which is linked to the evolutionary tracks and colours, may
be biased high at supra-solar metallicities. This problem affects mainly the most 
metal-rich, early-type galaxies.
To account for this bias, we include models producing dust attenuations down to $A_z=
-0.1$ when computing the PDFs of stellar mass, stellar metallicity and age.\footnote{The difference in $r-i$ colour between two
old stellar populations of metallicities $2.5\,Z_\odot$ and $2\,Z_\odot$, when 
interpreted as a colour excess, corresponds to a $z$-band attenuation $A_z\sim0.1$.}
 
A comparison of our stellar mass estimates with those derived by \citet[][ which were
based on a preliminary version of the BC03 models and an early calibration of the SDSS
spectra]{kauf03a} for the galaxies in common between the two samples shows overall 
consistency, with a scatter of $\sim0.16$ dex. Our stellar mass estimates tend to be 
systematically larger than theirs by $\sim0.1$ dex. A difference of almost $+0.04$ dex can 
be attributed to the different prior used to generate the Monte Carlo library. The remaining 
offset is likely to originate from their exclusion of all models with $r-i$ colour redder than 
that observed, while we include models which imply $A_z$ `attenuations' down to $-0.1$. 

\section{The ages and metallicities of nearby galaxies}\label{results}

\begin{figure}
\centerline{\includegraphics[width=8truecm]{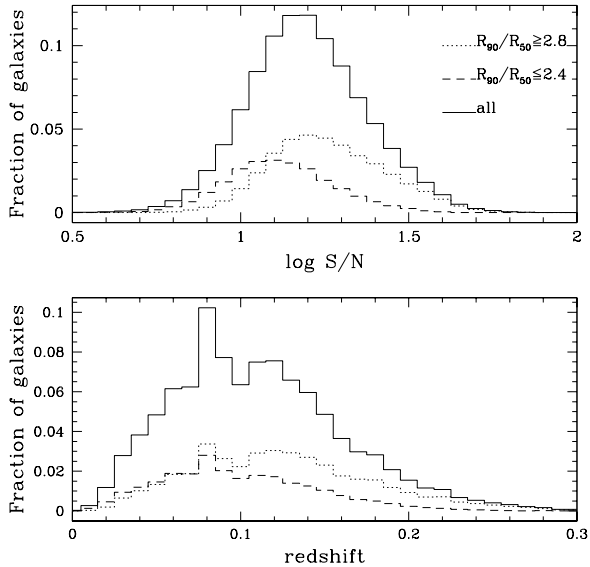}}
\caption{Distributions in median S/N per pixel (upper panel) and in redshift (bottom
panel) of the sample of 196,673 SDSS-DR2 galaxies (solid histograms). In each panel,
the dotted and dashed histograms show the corresponding distributions for the subsamples
of high-concentration ($C\geq2.8$), early-type galaxies and low-concentration ($C\leq2.4$),
late-type galaxies, respectively.}
\label{fig4}
\end{figure}

We now use the models described in Section \ref{approach} to estimate ages and
metallicities for a sample of 196,673 galaxies drawn from the SDSS-DR2 \citep[][see
also Section~\ref{spectra} above]{dr204}. Fig.~\ref{fig4} shows the distributions
in median S/N per pixel (upper panel) and in redshift (lower panel) of this sample.
Although the distribution in redshift extends up to 0.3, in what follows we discuss
results only for galaxies in the redshift range $0.005<z\leq0.22$. The lower limit is
intended to avoid redshifts for which deviations from the Hubble flow can be substantial,
but still allows us to include galaxies at very low luminosity (corresponding roughly
to a lower mass limit of $\sim 10^8 M_\odot$). The upper limit corresponds roughly to
the redshift at which a typical $10^{11}M_\odot$ galaxy is detected with median S/N
per pixel greater than 20. These cuts leave us with 175,128 galaxies.

In some applications below, it will be useful to distinguish between different 
morphological types of galaxies. This can be achieved on the basis of the `concentration
parameter' $C=R_{90}/R_{50}$, defined as the ratio of the radii enclosing 90 and 50 
percent of the Petrosian $r$-band luminosity of a galaxy. \cite{strateva01} and 
\cite{shimasaku01} have shown that the concentration parameter allows a rough 
classification into galaxy morphological types. \cite{strateva01} propose a cut at
$C=2.6$ to separate early- from late-type galaxies. To limit the contamination between
the two types,  here we define those galaxies with $C\geq2.8$ to be `early-type' 
and those galaxies with $C\leq2.4$ to be `late-type'. The dotted and dashed lines in 
Fig.~\ref{fig4} show separately the distributions in median S/N per pixel and
in redshift of the resulting subsamples of early- and late-type galaxies, respectively. 
Early-type galaxies generally have a higher median S/N per pixel and are detected
out to higher redshifts than late-type galaxies, as expected from their higher surface
brightnesses.

\begin{figure*}
\centerline{\includegraphics[width=10truecm]{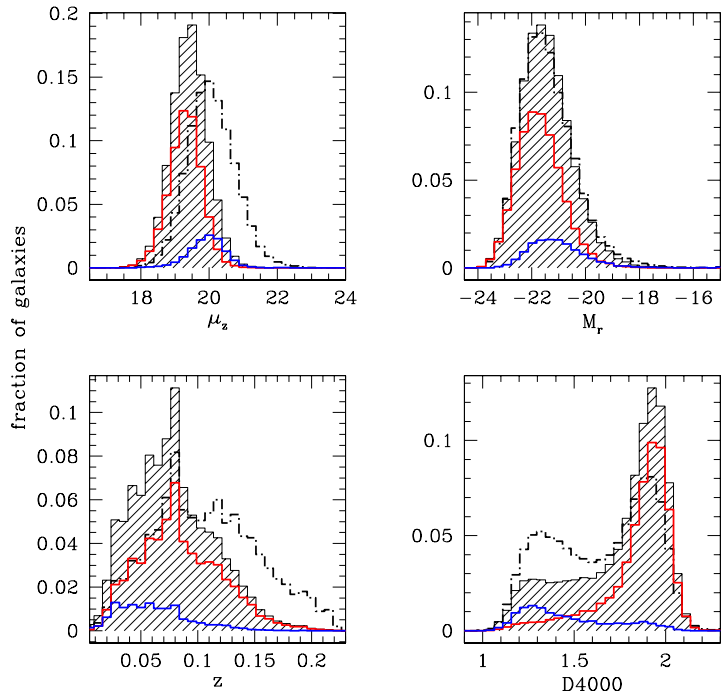}}
\caption{From top-left to bottom-right: distribution in $z$-band surface brightness, 
$r$-band absolute Petrosian magnitude (k-corrected at $z=0$), redshift and 4000\AA-break 
strength for the galaxies in our sample. The dot-dashed line represents the distribution for 
the original sample of 175,128 galaxies in the redshift range $0.005<z\leq0.22$. The shaded 
histogram describes the distribution for the final sample of 44,254 high-S/N galaxies, while 
the red and blue histograms separate (high-S/N) early-type ($C\geq2.8$) and late-type ($C\leq2.4$) 
galaxies, respectively. (Note that the dot-dashed and the shaded histograms are both normalized 
to unit area.)}\label{fig5}
\end{figure*}

We have shown in Section~\ref{accuracy1} and Fig.~\ref{fig3} the impact of the 
signal-to-noise ratio on our ability to derive reliable constraints on age and, 
especially, metallicity. In the remainder of this paper we focus 
on the subsample of 44,254 galaxies with median S/N per pixel greater than 20, for which 
the constraints on metallicity are the most accurate. The properties of these high-S/N 
galaxies are summarized in Fig.~\ref{fig5} where we show their distribution (shaded histogram) 
in $z$-band surface brightness (averaged within the Petrosian $R_{50}$ radius), Petrosian $r$-band 
absolute magnitude, redshift and \dn, compared to the distributions for the original sample 
(dot-dashed line). The red and blue lines distinguish early-type from late-type (high-S/N) galaxies. 
The S/N requirement excludes about 75 percent of our sample. As can be seen from 
Fig.~\ref{fig5}, this biases the sample toward higher-surface brightness, more concentrated, 
lower-redshift galaxies. There is not a bias instead in the luminosity distribution. 
In the redshift range $0.005 < z \leq 0.22$, high-S/N galaxies typically 
have $z$-band surface brightnesses brighter than 21 mag~arcsec$^{-2}$. They lie primarily at 
redshifts less than 0.15, where they account for about 40 percent of the galaxies. They also 
represent roughly 40 percent of the high-concentration sample but only 10 percent of the 
low-concentration sample. Similarly, they constitute about 30 percent of the galaxies with 
$\dn>1.6$ and 15 percent of those with $\dn<1.6$. Our selection in S/N therefore excludes 
a substantial fraction of diffuse systems, with potentially subsolar 
metallicities. We will turn our attention to these systems in the third paper of this series. 
Their omission here does not alter significantly the results presented in the next sections. 
In particular, we have checked that the median ages and metallicities found for galaxies with 
stellar masses less than $\sim 2\times 10^9M_\odot$ in Sections~\ref{res2} and \ref{agez} below 
would change by at most 0.2 dex if we included galaxies with low S/N in our sample.

\subsection{Age, metallicity and mass distributions}\label{res1}
\begin{figure*}
\centerline{\includegraphics[width=10truecm]{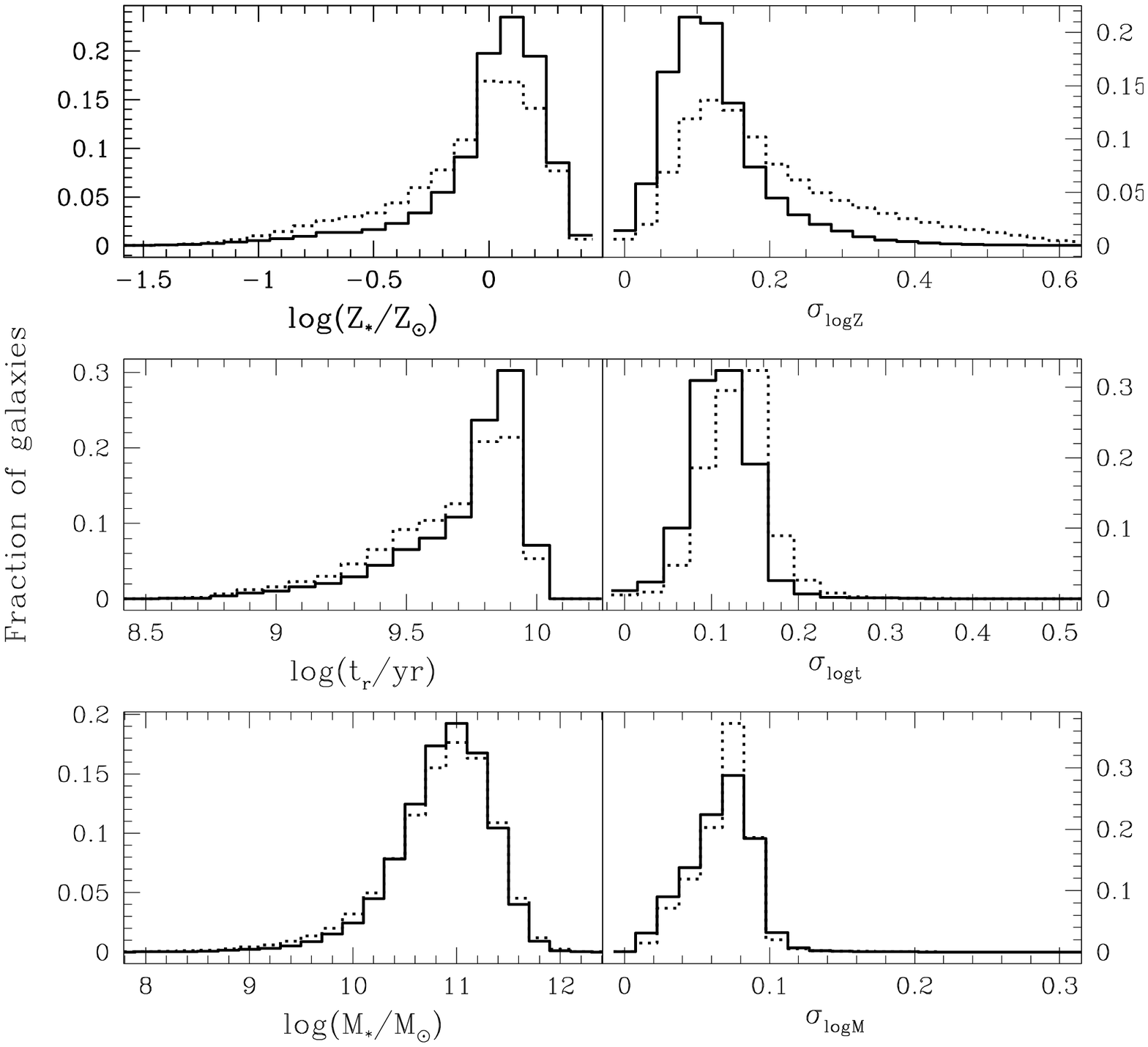}}
\caption{Distributions of the median-likelihood estimates of stellar metallicity 
(top-left panel), age (middle-left panel) and stellar mass (bottom-left panel) for 
44,254 SDSS-DR2 galaxies with redshifts in the range $0.005< z\leq 0.22$ and median 
S/N per pixel greater than 20 (solid histograms). The right-hand panels show the 
corresponding distributions of errors, computed as one half the 68 percent confidence 
ranges in the estimates of $\log(Z/Z_\odot)$, $\log(t_r/{\rm yr})$ and $\log(M_\ast/M_\odot)$. 
The dotted histograms are for the original sample of 175,128 galaxies (all histograms are 
normalized to unit area).}
\label{fig6}
\end{figure*}

We present here the distributions in age, metallicity and stellar mass of the 44,254
galaxies in our final sample. We recall that we estimate $r$-band light-weighted ages
but assume that all stars in a given galaxy have a single metallicity (interpreted as
the optical light-weighted metallicity). The left-hand panels of 
Fig.~\ref{fig6} show, from top to bottom, the distributions of the 
median-likelihood estimates of stellar metallicity, age and stellar mass for the galaxies
in our sample (solid histograms). The distribution in metallicity peaks around
$\log(Z/Z_\odot)\sim0.1$, the number of galaxies with sub-solar metallicities decreasing
smoothly down to 10 percent of solar. The distribution in age indicates that most of the
galaxies have fairly old mean stellar populations, with $t_r\sim7-8$ Gyr. The distribution
in stellar mass reveals that our sample is dominated by relatively massive galaxies, with
$M_\ast\sim10^{11} M_\odot$, while only a small fraction of the sample has masses below
$10^{10} M_\odot$. 

The right-hand panels of Fig. \ref{fig6} show the distributions of one half the
68 percent confidence range in the estimates of $\log(Z/Z_\odot)$, $\log(t_r/{\rm yr})$
and $\log(M_\ast/M_\odot)$. Stellar mass is the best constrained parameter, with a typical 
random uncertainty of $\sim0.08$ dex. Almost all the galaxies in the sample have a stellar mass 
estimate with an uncertainty less than 0.1 dex. Also, the ages are constrained within 
0.2 dex for almost all the galaxies, the typical uncertainty being $\sim0.12$ dex. 
Similarly the average error on stellar metallicity is 0.12 dex, but the distribution extends 
to larger errors ($\sim0.3$) than for light-weighted age. 

As a comparison, the dotted histograms show the distributions in the derived parameters and 
the associated uncertainties for the original sample of 175,128 galaxies. There is a 
significant tail of galaxies with very broad stellar metallicity PDFs, resulting in uncertainties 
greater than 0.25 dex. This further demonstrates that the uncertainties in metallicity estimates 
depend sensitively on the S/N in the observed spectra, as discussed in Section~\ref{accuracy}.
Also, the distribution in stellar metallicity shows that with our S/N cut we have excluded 
preferentially galaxies with subsolar metallicity. High-metallicity galaxies tend 
to be associated with small errors and low-metallicity galaxies with large errors (presumably 
because of their weaker absorption lines). 
Including low-S/N galaxies also increases the average error on light-weighted age to $\sim0.15$ dex, 
but the trend of increasing error with decreasing age is less strong than the analogous trend 
for metallicity.
 
\subsection{Relations between age, stellar metallicity, stellar mass and gas-phase 
metallicity}\label{res2}
It is interesting to examine how age, metallicity and stellar mass (and the errors on 
these quantities) are distributed among galaxies with different star formation histories.
As shown by \cite{kauf03a}, galaxies with different star formation histories populate 
different areas of a diagram defined by \dn\ and the strength of a H-Balmer absorption
line (they illustrated this result for $\rm H\delta_A$). Galaxies with smooth star 
formation histories form a sequence extending from actively star-forming galaxies 
(corresponding to small \dn\ and strong H-Balmer absorption) to quiescent early-type
galaxies (corresponding to large \dn\ and weak H-Balmer absorption). Galaxies which
experienced recent bursts of star formation exhibit the strongest H-Balmer absorption 
at fixed \dn.

\begin{figure*}
\centerline{\includegraphics[width=12truecm,height=14truecm]{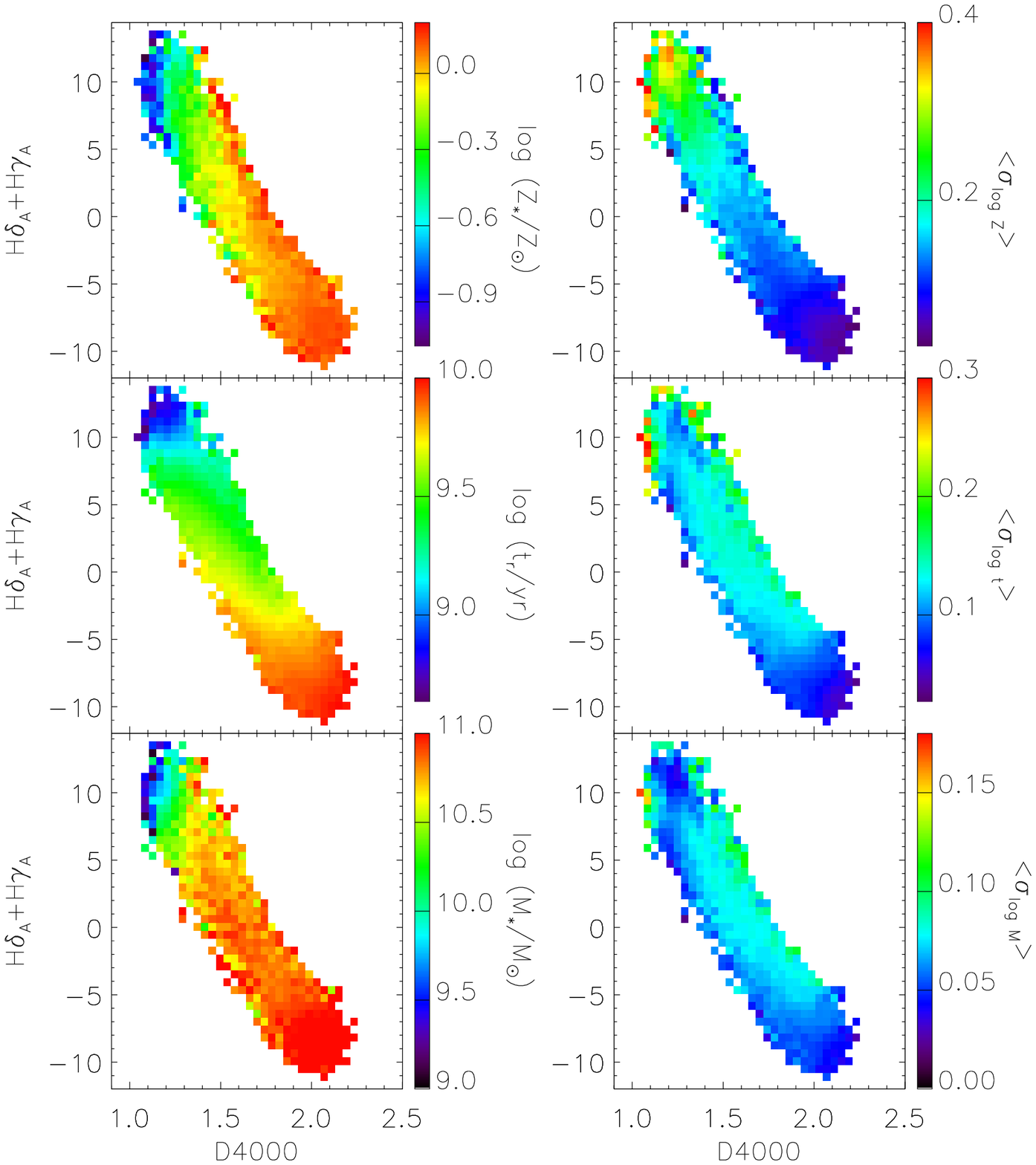}}
\caption{Physical parameters of 44,254 SDSS-DR2 galaxies with median S/N per pixel greater
than 20 as a function of position in the \hdg\ versus \dn\ diagram. In the left-hand 
panels, the diagram has been binned and colour-coded in order to reflect the average 
stellar metallicity (top), age (middle) and stellar mass (bottom) of the galaxies falling
into each bin. In the right-hand panels, the diagram has been binned and colour-coded
to reflect the average uncertainty (corresponding to one half the 68 percent confidence
interval) in the derived stellar metallicities (top), ages (middle) and stellar masses 
(bottom) of the galaxies falling into each bin. The widths of the bins in \dn\ and \hdg\
correspond roughly to the mean observational errors in these quantities (0.04 and 0.4, 
respectively).}
\label{fig7} 
\end{figure*}

We now explore how the 44,254 high-S/N galaxies in our sample populate such a 
diagram as a function of their physical parameters derived in Section~\ref{res1} above.
We consider here for consistency the diagram defined by \dn\ and the \hdg\ absorption
index. In the left-hand panels of Fig.~\ref{fig7}, we have binned and colour-coded 
this diagram in order to reflect, from top to bottom, the average stellar metallicity,
the average age and the average stellar mass of the galaxies falling into each bin. The
widths of the bins in \dn\ and \hdg\ correspond roughly to the mean observational errors
in the two quantities for galaxies with median S/N per pixel greater than 20 (0.04 and 
0.4, respectively). 

The bottom-left panel of Fig.~\ref{fig7} shows that stellar mass increases with
\dn\ along the sequence populated by SDSS galaxies in the \hdg\ versus \dn\ diagram. 
The average mass of galaxies with $\dn\la1.2$ is less than $10^{10}M_\odot$,
while that of galaxies with $\dn\ga1.9$ reaches $10^{11}M_\odot$. There is a smooth 
transition between these two regimes in the central region of the diagram. This confirms
the trends in \dn\ and H$\delta_A$ as functions of stellar mass observed by \cite{kauf03b} 
(see their fig. 1). Both stellar metallicity and age also appear to increase
with increasing \dn\ (upper- and middle-left panels of Fig.~\ref{fig7}). The variation
in age is very smooth. In the intermediate regime, age appears to decrease with 
increasing \hdg\ at almost fixed \dn. The variation in metallicity is more noisy, partly
because of the larger errors associated with metallicity estimates, especially for low
\dn\ (see Section~\ref{res1}).  Near the middle of the galaxy sequence, i.e., for $1.4
\la\dn\la 1.8$, stronger H-Balmer absorption appears to be associated to not only younger
but also more metal-rich stellar populations. This is consistent with the idea that these
galaxies could have experienced a burst of metal-enriched star formation about 1--2 Gyr 
ago.\footnote{At fixed $\dn \sim1.5$, this trend in age and metallicity is not consistent
with (and more pronounced than) the age-metallicity degeneracy described in 
Section~\ref{agez}.}

We have checked that the above trends in stellar metallicity, age and stellar mass do 
not change significantly if we include galaxies with lower S/N values. The main effect 
is to increase the scatter in these physical parameters along the \hdg\ versus \dn\ 
sequence, mainly because of the larger observational errors in these two indices. 
Moreover, the average S/N for the full sample does not vary significantly along the 
relation. For these reasons, we believe that our results are not strongly biased by 
our cut in S/N.

In the right-hand panels of Fig.~\ref{fig7}, we have colour-coded the \hdg\ versus \dn\
diagram to reflect the average uncertainties in the determinations of stellar metallicity,
age and stellar mass (from top to bottom) for the galaxies falling into each bin. These
diagrams illustrate how the tightness of the constraints derived on the various physical 
parameters depends on the strengths of \hdg\ and \dn.  The errors in all three parameters
tend to be larger at lower \dn, the trend being especially strong for stellar metallicity.
We emphasize that the right-hand panels of Fig.~\ref{fig7} do {\em not} reflect the rms
scatter in the various physical parameters, but rather the average uncertainties 
associated with determinations of these parameters. For reference, the rms scatter is
typically comparable to (or smaller than) the average uncertainty for stellar metallicity
and age, but always larger than the average uncertainty for stellar mass.

\begin{figure*}
\centerline{\includegraphics[width=15truecm]{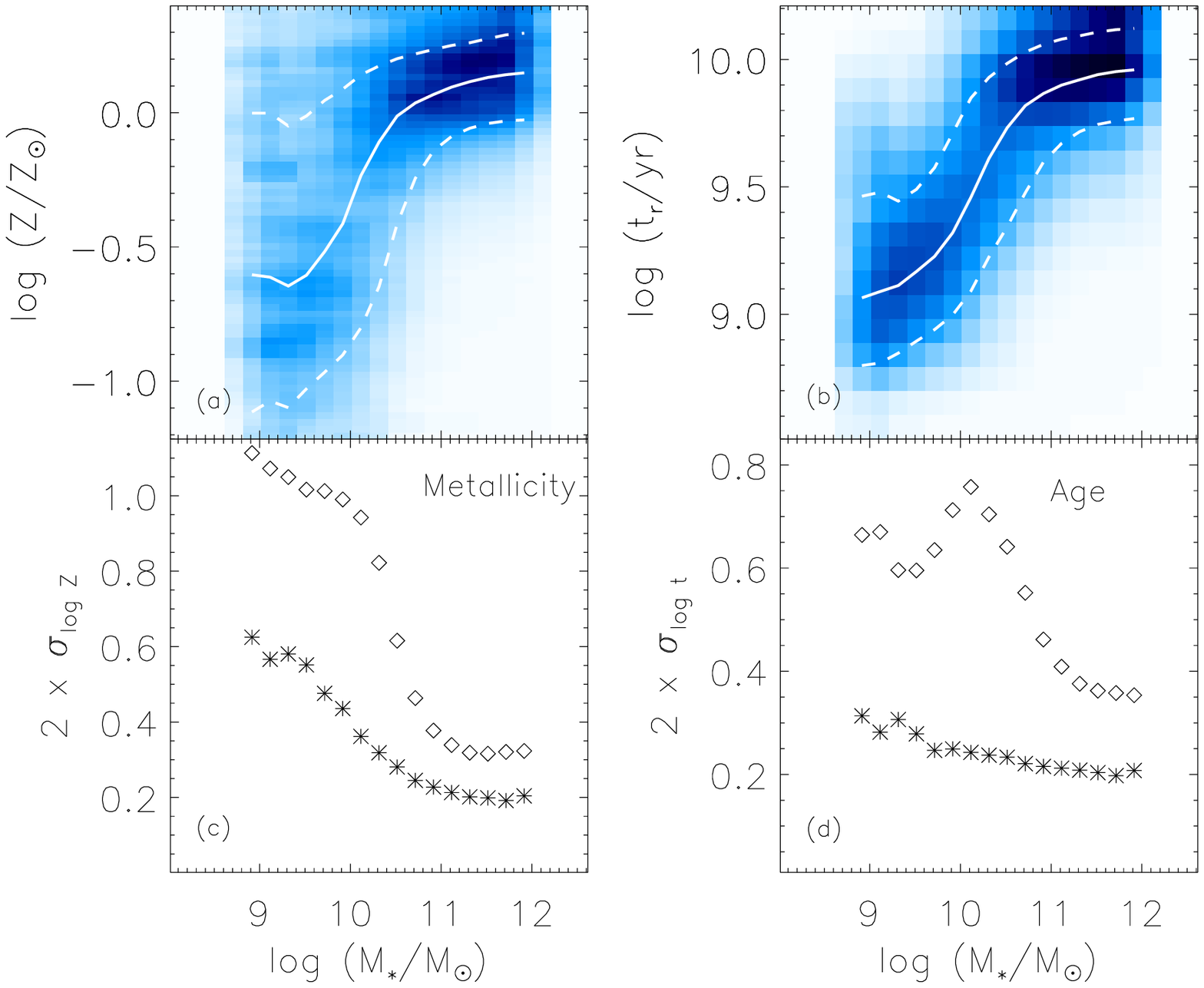}}
\caption{Panels a and b: conditional distribution of stellar metallicity and age 
as a function of stellar mass for 44,254 SDSS-DR2 galaxies with median
S/N per pixel greater than 20. Each distribution was obtained by coadding the normalized
2-D likelihood distributions of the desired parameter (stellar metallicity or age) and 
stellar mass for all the galaxies and then re-normalizing along the y-axis in bins of 
stellar mass (of width 0.2 dex). The solid line indicates the median of the final 
conditional distribution, and the dashed lines the $16^{\rm th}$ and $84^{\rm th}$ 
percentiles. Panels c and d: comparison of the scatter in the metallicity and age 
distributions with the mean uncertainty of the metallicity and
age estimates as a function of stellar mass. The diamonds show the difference in 
$\log (Z/Z_\odot)$ or $\log (t_r/{\rm yr})$ between the $\rm 16^{th}$ and $\rm 
84^{th}$ percentiles of the conditional distributions of panels a and b, while
the stars represent the mean 68 percent confidence range in $\log (Z/Z_\odot)$ or 
$\log (t_r/{\rm yr})$ for the galaxies in each stellar mass bin.}\label{fig8}
\end{figure*}

The results of Fig. \ref{fig7} suggest that both stellar metallicity and age correlate 
with stellar mass. We investigate this further in Fig.~\ref{fig8}, where we show the 
distributions of metallicity (panel a) and age (panel b) as function of
stellar mass for the 44,254 high-S/N galaxies in our sample. Rather than assigning each
galaxy its median-likelihood estimate of each of the three parameters, we keep here the
whole information contained in the PDFs. In Fig.~\ref{fig8}a, 
the likelihood distribution of stellar metallicity as a function of stellar mass was 
obtained by coadding the normalized 2-D likelihood distributions of metallicity and 
stellar mass for all the galaxies and then re-normalizing along the metallicity axis 
in bins of stellar mass (we adopted a bin width of 0.2 dex, roughly comparable to the 
68 percent confidence range in stellar mass estimates). The solid line indicates the 
median of the final conditional distribution, and the dashed lines the $16^{\rm th}$ 
and $84^{\rm th}$ percentiles. The percentiles of the metallicity and age distributions 
at fixed stellar mass are provided in Table~\ref{ztm_tab}.

The median shows that metallicity increases with stellar mass, from roughly 20 percent 
of solar for galaxies with masses below $10^{10}M_\odot$ to about 1.4 times solar for
those with masses above $10^{11}M_\odot$. In between, stellar metallicity increases
rapidly with stellar mass, the trend becoming shallower at $M_\ast\ga10^{10.5}
M_\odot$. Fig.~\ref{fig8}b shows the analogous conditional 
distribution of age as a function of stellar mass. The median ($r$-band light-weighted)
age increases from roughly 1\,Gyr for galaxies with masses around $10^{9}M_\odot$ to about
6.5\,Gyr for those with masses around $10^{10.5}M_\odot$. For more massive galaxies, the
median age continues to increase, but only by about 0.2 dex ($\sim1.6$ Gyr) over one order
of magnitude in stellar mass. 

\begin{table}
\caption{Median (P50) and percentiles (P16, P84) of the distributions in stellar metallicity 
and age as a function of stellar mass (Fig.~\ref{fig8}a,b).}\label{ztm_tab}
\centering
\begin{tabular}{|l|ccc|ccc|}
\multicolumn{3}{|c|}{}\\
\noalign{\smallskip}
\hline
 & \multicolumn{3}{c}{$\log(Z/Z_\odot)$} & \multicolumn{3}{c}{$\log(t_r/yr)$}\\
$\log(M_\ast/M_\odot)$ & P50 & P16 & P84 & P50 & P16 & P84\\
\noalign{\smallskip}
\hline
\noalign{\smallskip}			    
   8.91  &   -0.60  &	-1.11  &   -0.00 & 9.06  &    8.80  &	 9.46 \\
   9.11  &   -0.61  &	-1.07  &   -0.00 & 9.09  &    8.81  &	 9.48 \\
   9.31  &   -0.65  &	-1.10  &   -0.05 & 9.11  &    8.85  &	 9.44 \\
   9.51  &   -0.61  &	-1.03  &   -0.01 & 9.17  &    8.89  &	 9.49 \\
   9.72  &   -0.52  &	-0.97  &    0.05 & 9.23  &    8.94  &	 9.57 \\
   9.91  &   -0.41  &	-0.90  &    0.09 & 9.32  &    9.00  &	 9.71 \\
  10.11  &   -0.23  &	-0.80  &    0.14 & 9.46  &    9.09  &	 9.85 \\
  10.31  &   -0.11  &	-0.65  &    0.17 & 9.61  &    9.23  &	 9.93 \\
  10.51  &   -0.01  &	-0.41  &    0.20 & 9.73  &    9.34  &	 9.98 \\
  10.72  &    0.04  &	-0.24  &    0.22 & 9.82  &    9.48  &	10.03 \\
  10.91  &    0.07  &	-0.14  &    0.24 & 9.87  &    9.60  &	10.06 \\
  11.11  &    0.10  &	-0.09  &    0.25 & 9.90  &    9.67  &	10.08 \\
  11.31  &    0.12  &	-0.06  &    0.26 & 9.92  &    9.72  &	10.09 \\
  11.51  &    0.13  &	-0.04  &    0.28 & 9.94  &    9.75  &	10.11 \\
  11.72  &    0.14  &	-0.03  &    0.29 & 9.95  &    9.76  &	10.12 \\
  11.91  &    0.15  &	-0.03  &    0.30 & 9.96  &    9.77  &	10.12 \\
\noalign{\smallskip}
\hline
\end{tabular}
\end{table}

Although the median metallicity and age both increase with increasing stellar mass, the
probability levels (indicated by the colour) and the 68 percent confidence ranges reveal 
broad distributions in the estimates of both parameters. We now explore whether the 
larger scatter in both distributions at smaller masses can be accounted for by the 
larger uncertainties in metallicity and age estimates for more metal-poor and younger
galaxies, or whether it is indicative of an intrinsically broad distribution in 
metallicity and age for low-mass galaxies. In the bottom panels of Fig.~\ref{fig8}, we compare 
the scatter in these relations with the uncertainties in the estimates of metallicity
(panel c) and age (panel d) at fixed stellar mass. In each panel, the
diamonds show the difference between the $\rm 16^{th}$ and $\rm 84^{th}$ percentiles 
of the conditional distribution shown above as a function of stellar mass.
The stars show the mean 68 percent confidence range in the individual estimates of 
metallicity or age for the galaxies falling into each stellar-mass bin.\footnote{The 
stars in Fig.~\ref{fig8}c,d are not computed using the joint likelihood 
distribution, but rather by adopting the median of the $M_\ast$ PDF as our estimate of 
stellar mass. The comparison of the stars with the diamonds is consistent, given that the
width of the stellar mass bins is comparable with the error in stellar mass estimates.} 
Fig.~\ref{fig8}c shows that the scatter in the mass-metallicity relation 
is always larger than the uncertainties in the metallicity estimates, 
by at least 0.1 dex, and is largest (compared to the error) at about $10^{10}M_\odot$. 
Similarly, the scatter in the mass-age relation (Fig.~\ref{fig8}d) is always larger than the 
uncertainties in the age estimates and is also largest at about $10^{10}M_\odot$. Fig.~\ref{fig8} 
clearly indicates that more massive galaxies are older and more 
metal-rich, while less massive galaxies are younger and more metal-poor. Metallicity 
and age, however, are not uniquely determined by stellar mass. There is an intrinsic 
scatter in both parameters, which is particularly evident at intermediate stellar
masses. The results of Section \ref{agez} below indicate that part of this intrinsic
scatter can be accounted for by differences in galaxy morphology, the intermediate-mass 
regime corresponding to the transition between disc-dominated and bulge-dominated 
galaxies. However, a significant scatter persists even when considering the two classes of
galaxies separately.

\begin{figure}
\centerline{\includegraphics[width=9truecm]{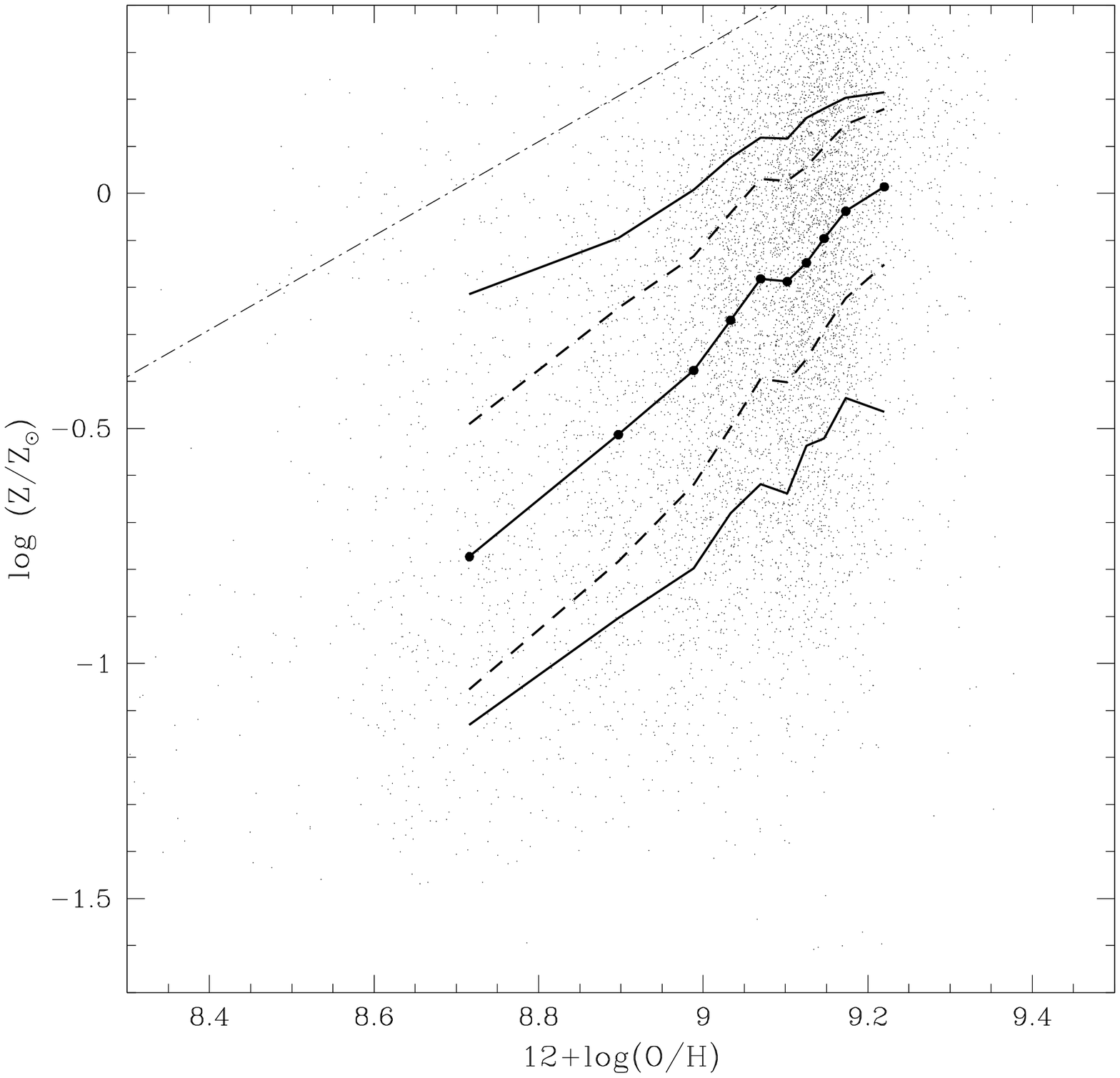}}
\caption{Stellar metallicity estimates (present work) plotted against gas-phase 
oxygen abundance 12+log(O/H) \citep{christy04} for 7462 high-S/N SDSS-DR2 galaxies
for which both measures are available. The small dots show the median-likelihood estimates
of the two parameters for each galaxy. The larger points (joined by a solid line)
show the median stellar metallicity in bins of 12+log(O/H) (each bin containing 
$\sim300$ galaxies), while the outer solid lines show the corresponding $16^{\rm th}$ and 
$84^{\rm th}$ percentiles. The dashed lines indicate the mean 68 percent confidence range
in the stellar metallicity estimates for the galaxies in each bin of 12+log(O/H). Solar
metallicity is 8.69 in these units \citep{allende01}. The dot-dashed line represents the one-to-one 
relation, showing that stellar metallicity is always lower than gas-phase metallicity.}\label{fig9}
\end{figure}

\cite{christy04} have shown that the emission-line galaxies in the DR2 sample exhibit 
a tight relation between stellar mass and gas-phase oxygen abundance, as determined from
their nebular spectra. The gas-phase oxygen abundance increases steadily from low to high
stellar masses, and then gradually flattens around masses of $10^{10.5}M_\odot$. There 
is a striking similarity between the relation \cite{christy04} found for star 
forming galaxies and the relation found in Fig.~\ref{fig8} above between stellar 
metallicity and stellar mass, which includes both star-forming and quiescent galaxies. 
However, the dispersion in the stellar metallicity versus stellar mass relation is much
larger than that in the relation between gas-phase oxygen abundance and stellar mass 
($\pm0.1$ dex). This may partly reflect the larger errors associated with stellar 
metallicity determinations. 

In Fig.~\ref{fig9}, we plot stellar metallicity against gas-phase oxygen abundance, 
12+log(O/H), for the 7462 high-S/N galaxies of our sample for which both measures are 
available. We note that gas-phase metallicity is best determined for star-forming 
galaxies, whereas stellar metallicity is best determined for early-type galaxies. 
The small dots show the median-likelihood estimates of the stellar and gas-phase 
metallicities for each galaxy. The larger points indicate the median stellar metallicity
in bins of 12+log(O/H), while the outer solid lines show the corresponding $16^{\rm
th}$ and $84^{\rm th}$ percentiles. There is a relation between the two parameters 
with approximately unit slope. However, the stellar metallicity is generally lower than the 
gas-phase metallicity (by $\sim0.5$dex), as demonstrated by the one-to-one relation 
(dot-dashed line). This is expected since the nebular metallicity traces the metallicity 
of the last generations of stars to form, whereas the stellar metallicity represents an 
average over the entire star formation history of the galaxy. Notably, the large scatter 
in stellar metallicity at fixed gas-phase oxygen abundance (solid lines) cannot be solely 
accounted for by the errors in stellar metallicity estimates (indicated by the dashed lines). 

\begin{figure*}
\centerline{\includegraphics[width=13truecm]{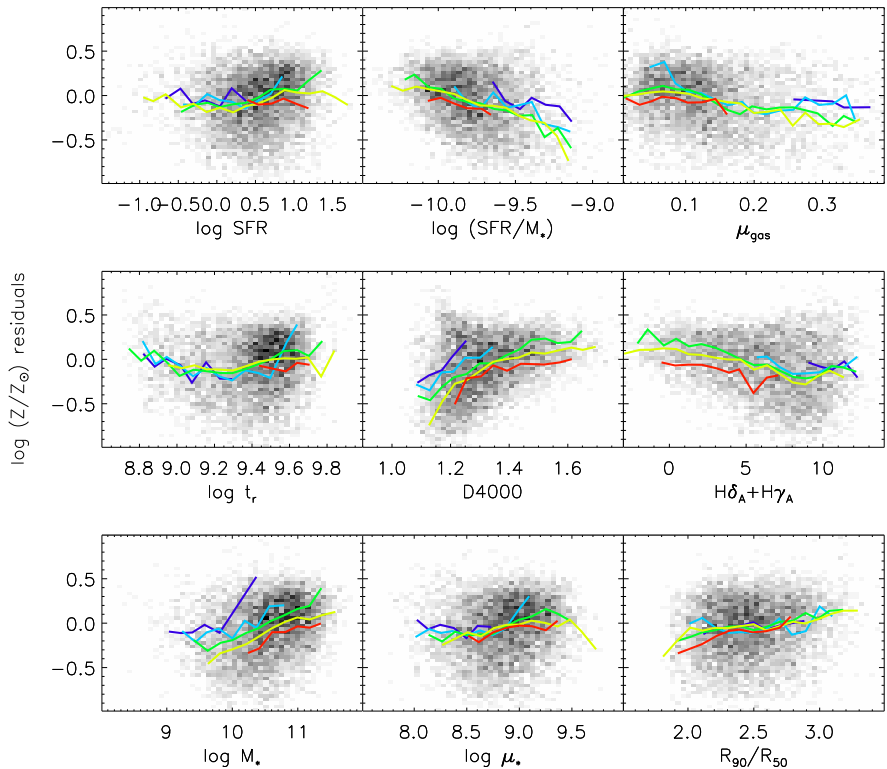}}
\caption{Residuals in $\log(Z/Z_\odot)$ with respect to a simple linear fit of the 
median relation between stellar metallicity and gas-phase oxygen abundance in
Fig.~\ref{fig9}, plotted against various galaxy properties. From left to right, 
top raw: star formation rate (in units of $M_\odot\rm yr^{-1}$), specific star
formation rate (in units of $\rm yr^{-1}$) and gas mass fraction. Middle raw: 
$r$-band light-weighted age (in units of yr), \dn\ and \hdg\ index strengths. 
Bottom raw: total stellar mass (in units of $M_\odot$), surface mass density 
within the Petrosian $r$-band $R_{50}$ radius (in units of $M_\odot\rm  kpc^{-2}$)
and $r$-band concentration parameter. In each panel, curves of different colors
indicate the median relations in different bins of gas-phase oxygen abundance, 
increasing from $8.7\pm0.075$ (blue) to $9.3\pm0.075$ (red).}\label{fig10}
\end{figure*}

To investigate the origin of this scatter, we compute the residuals in $\log(Z/Z_\odot)$
with respect to a simple linear fit of the median relation (large dots) in Fig.~\ref{fig9}.
The residuals are plotted in Fig.~\ref{fig10} as a function of various galaxy properties,
including the star formation rate (from \citealt{jarle03}) and an indirect estimate of
the gas mass fraction inferred from the star formation rate (see eq.~5 of 
\citealt{christy04}). In each panel, we indicate the median trend of the residuals as a
function of the property under consideration for galaxies in five bins of gas-phase 
oxygen abundance, from $8.70\pm0.075$ (blue) to $9.3\pm0.075$ (red). The $\log(Z/Z_\odot)$
residuals show positive correlations with stellar mass, surface mass density, concentration
parameter and \dn\ and negative correlations with specific star formation rate, gas mass
fraction and \hdg. These various trends are almost independent of gas-phase 
oxygen abundance.\footnote{We checked that the trends of $\log(Z/Z_\odot)$ residuals 
with the strengths of \dn\ and \hdg\ are not caused by any bias in our estimates of 
stellar metallicity (see Section~\ref{accuracy2}).} They imply that, at fixed 12+log(O/H),
the stars are most metal-rich in massive, high-concentration galaxies, which have
exhausted most of their gas and form stars at a lower rate than in the past. The existence
of a substantial range of stellar metallicities at fixed gas-phase oxygen abundance in 
Fig.~\ref{fig9} further suggests that the galaxies in our sample are not well approximated by
`closed-box' systems, for which stellar and interstellar metallicities should be tightly
related \citep[e.g.,][]{tinsley80}. This indicates that gas ejection and/or accretion may be 
important factors in galaxy chemical evolution.

\subsection{Age versus metallicity}\label{agez}

Most previous population synthesis studies of the age-metallicity relation for nearby
galaxies have focused on early-type galaxies \citep[e.g.][]{gonzalez93,worthey97c,bernardi98,
terlevich99,fcs99,trager00b}. 
A few studies have also included the bulges of spiral galaxies \citep{jablonka96,GGJ99,tf01,
proctor02}. Our determination, with a new method, of the ages and metallicities of a large sample 
of nearby galaxies spanning a wide range of star formation activities allows us to re-assess the 
relation between these two physical parameters.

\begin{figure*}
\centerline{\includegraphics[width=10truecm]{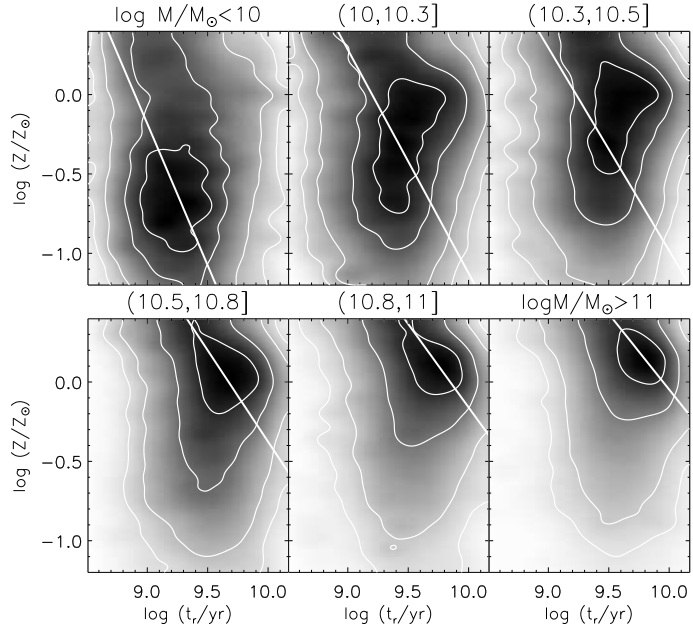}} 
\caption{Two-dimensional distribution of stellar metallicity and age in 6 bins
of stellar mass (as indicated) for the subsample of 5616 late-type ($C
\leq2.4$) SDSS-DR2 galaxies with median S/N per pixel greater than 20. Each 
distribution was obtained by coadding the normalized joint likelihood distributions 
of age and metallicity for individual galaxies. The grey levels indicate the probability 
associated with each value of metallicity and age in a square-root scale, while 
the contours enclose 26, 68, 95 and 99 percent of the total probability. The straight
line (which intersects the peak of the probability distribution) indicates the mean 
slope of the age-metallicity degeneracy for the galaxies falling into each stellar-mass bin.}
\label{fig11}
\end{figure*}

\begin{figure*}
\centerline{\includegraphics[width=10truecm]{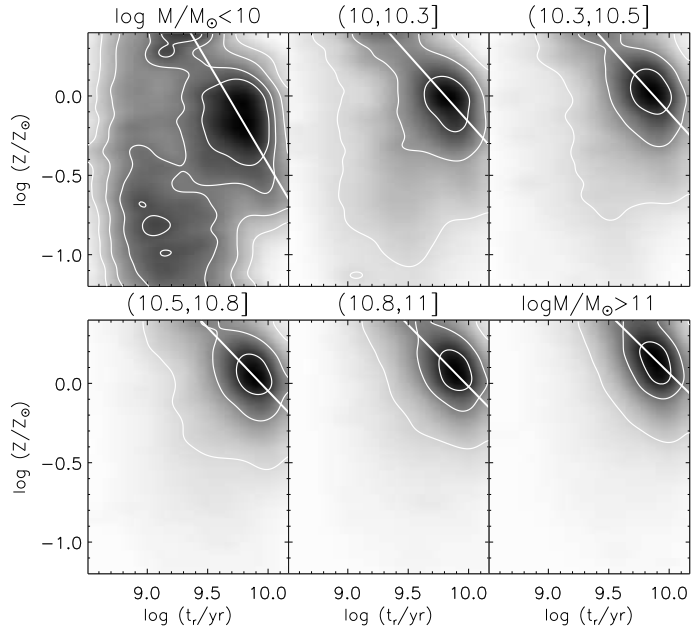}} 
\caption{Same as Fig.~\ref{fig11}, but for the subsample of 26,003 early-type ($C
\geq2.8$) SDSS-DR2 galaxies with median S/N per pixel greater than 20.}
\label{fig12}
\end{figure*}

We separate here high-concentration, bulge-dominated early-type galaxies and 
low-concentration, disk-dominated late-type galaxies as described above to understand
whether different conclusions can be drawn for different types of galaxies. We have
shown in Fig.~\ref{fig8} that both metallicity and age increase with stellar mass.
We may therefore expect that the distribution of galaxies over metallicity and age
will change when different masses are considered. This mass dependence may also be 
different for different types of galaxies. For this reason, we divide each morphological
subsample into six bins of stellar mass. Figs~\ref{fig11} and \ref{fig12}
show the age-metallicity relation in these different stellar-mass bins for late-type and
early-type galaxies, respectively. The stellar mass increases from $\log (M/M_\odot)
\leq10$ in the top-left panel to $\log(M/M_\odot)>11$ in the bottom-right panel. These
relations were obtained by simply coadding the normalized joint likelihood distributions
of age and metallicity for individual galaxies, without further normalization. The grey
levels indicate the probability associated with each value of metallicity and age in a 
square-root scale, while the contours enclose 26, 68, 95 and 99 percent of the total 
probability. 

Fig.~\ref{fig11} shows that the ages and metallicities of low-concentration, 
late-type galaxies change significantly as a function of stellar mass. Galaxies 
less massive than $10^{10} M_\odot$ have typically low metallicities ($Z\approx
10^{-0.6}Z_\odot$) and young ages ($0.8\la t_r/{\rm Gyr}\la2.5$). At 
intermediate stellar masses, i.e. for $10^{10}\la M_\ast/M_\odot\la10^{10.5}$, the 
distribution in metallicity becomes broader and shifts gradually to higher 
metallicities. The overall spread in age is roughly 0.8 dex, 
and the distribution shifts slightly to older ages. Late-type galaxies more massive 
than $10^{10.5} M_\odot$ have predominantly high metallicities ($Z\ga Z_\odot$) and 
old ages ($3.2\la t_r/{\rm Gyr}\la10$). The fraction of young metal-poor 
galaxies nearly vanishes at the highest masses.

Fig.~\ref{fig12} shows that the ages and metallicities of high-concentration,
early-type galaxies depend much more weakly on stellar mass than those of 
low-concentration, late-type galaxies. The gradual increase of metallicity with
stellar mass is still appreciable for early-type galaxies. Galaxies less massive
than $10^{10} M_\odot$ have metallicities typically in the range $10^{-0.4}$--$10^{
0.1}Z_\odot$, while galaxies more massive than $10^{11} M_\odot$ have metallicities
typically above solar. The age range does not exhibit
any significant variation with stellar mass, except for a population 
of young metal-poor galaxies in the lowest-mass bin, which does not show up at higher
masses. At low stellar masses, we also notice a small fraction ($\sim1$ percent) of
galaxies clumping at much higher metallicities and younger ages than the bulk of the 
early-type galaxy sample. Detailed inspection reveals that these galaxies are photometric
outliers, for which we are uncertain of the accuracy of our fits. Follow-up 
observations would be required to understand whether the unusual metallicities 
and ages of these galaxies are consistent with a substantial recent episode of 
metal-enriched star formation. 

We note that the typical light-weighted age for the massive 
early-type galaxies in our sample is $\sim$8 Gyr, which is somewhat younger than the ages 
usually quoted for (cluster) ellipticals. Differences can arise from differences in the 
adopted methods. In particular, the ages usually quoted in the literature are derived by 
comparing the observed distributions in spectral indices with simple stellar population 
predictions. What we quote here is a proper light-weighted age, calculated by weighting 
the age of each generation of stars by their luminosity along the entire SFH of each model 
galaxy. Nevertheless, the relatively young ages we find could reflect an environmental 
dependence, since the SDSS sample is composed predominantly of galaxies in lower density 
regions than those used in most studies of early-type galaxies. Indeed, several studies have 
reported an age difference between early-type galaxies in clusters and in low-density regions 
from $\sim$1.2 Gyr \citep{bernardi98} up to $2-3$ Gyr \citep{kunt02,thomas05}. We will discuss 
the properties of the early-type galaxies in our sample in more detail in the second paper of 
this series.

A comparison of Figs~\ref{fig11} and \ref{fig12} suggests that the ages
and metallicities of late- and early-type galaxies differ primarily at the low-mass
end. Low-mass late-type galaxies tend to be younger and more metal-poor than their 
early-type counterparts (by about 0.6~dex in both age and metallicity). In contrast,
massive late-type galaxies tend to have old ages and high metallicities similar to
those of their early-type counterparts, even though there is a tail of young ($t_r\la
10^{9.5}$ yr), metal-poor ($Z\approx10^{-0.5}Z_\odot$) late-type galaxies with masses
greater than $10^{11} M_\odot$ in our sample. The high metallicities of most massive
late-type galaxies might reflect the influence of their metal-rich bulges. This may
be enhanced by the fact that the SDSS spectra sample only the inner regions of the 
galaxies (see Section \ref{aperture} below). 

Figs~\ref{fig11} and \ref{fig12} also indicate that the relation between stellar metallicity 
and stellar mass discussed in Section~\ref{res2} holds for both late-type and early-type galaxies. 
In contrast, the relation between age and stellar mass is much more pronounced for late-type galaxies, 
which cover a significant range in age ($\sim9$ Gyr). Early-type galaxies do not show  
such a clear trend, at least above $10^{10.5} M_\odot$. This is in agreement 
with previous findings of no significant relation between the ages of early-type galaxies and 
luminosity \citep{KD98,kunt98,tf01}. At fainter luminosities, however, there are indications 
of a spread to younger ages \citep[see also][]{worthey97,CR98}. We also find that the fraction 
of young early-type galaxies increases at lower stellar masses. This is true in particular below 
$\rm 10^{10} M_\odot$, reflecting a possible contamination by S0 galaxies \citep[e.g.][]{poggianti01b}. 
Finally, we note that for both early-type galaxies and high-mass late-type galaxies, the probability 
contours of Figs~\ref{fig11} and \ref{fig12} hint at an anticorrelation between age and stellar 
metallicity, at given mass. Other workers have pointed out this anticorrelation 
\citep[e.g.][]{wtf95,colless99,jorgensen99,rakos01,poggianti01a}, but it is difficult to assess 
its significance here because of the correlated errors on age and metallicity (see below). 

The dependence of the properties of SDSS galaxies on stellar mass has been addressed
in previous studies \citep[e.g. ][]{shen03,jarle03,christy04,kannappan04,salim05,jimenez05}. 
In particular, \cite{kauf03b} point out that, at masses around
$\sim 10^{10.5}M_\odot$, galaxies separate into two distinct classes of spectral
and structural properties. \cite{baldry04} have recently shown that this transition around 
$(2-3)\times10^{10} M_\odot$ can be seen directly from the colour-magnitude relation. Moreover, 
this mass range seems also to correspond to a shift in gas richness \citep{kannappan04}. It is 
interesting to note that, around the same value of stellar mass, there appears to be also a gradual 
transition in the ages and metallicities of the galaxies. This transition, which is particularly 
evident for late-type galaxies, causes the dispersion in age and metallicity to be highest for
galaxies of intermediate mass.


We have not mentioned so far the potential influence of the `age-metallicity degeneracy'
on the results of Figs~\ref{fig11} and \ref{fig12} (see Section~\ref{intro}).
This degeneracy, which arises from the competing effects of age and metallicity on
absorption-line strengths, can cause the confusion between old, metal-poor galaxies and
young, metal-rich ones \citep{worthey94}. To address this issue, we first evaluate the slope 
of the age-metallicity degeneracy from the two-dimensional likelihood distribution of 
each individual galaxy in our sample. This is achieved by computing the matrix of
second moments of the likelihood distribution (inside the 68 percent probability contour)
with respect to the median $\log Z$ and $\log t_r$ values. The direction of the eigenvector
corresponding to the largest eigenvalue of this matrix defines the slope $\Delta\log Z
/\Delta\log t_r$ of the age-metallicity degeneracy. In each panel of Figs~\ref{fig11}
and \ref{fig12}, the straight line indicates the mean $\Delta\log Z/\Delta\log t_r$
for the galaxies falling into the corresponding bin of stellar mass. The slopes are listed
in Table~\ref{slopes}.\footnote{These values are consistent with the results of 
\citet{worthey96}, which imply that the change in metallicity needed to compensate a change
in age at fixed index strength in the spectrum of an old stellar population should 
correspond to values of $\Delta\log Z/\Delta\log t_r$ ranging from $-1.7$ for age-sensitive
indices to $-0.2$ for metal-sensitive indices.} They tend to be steeper for 
low-concentration galaxies ($-1.78$ to $-0.95$) than for high-concentration ones ($-1.29$ 
to $-0.75$).
\begin{table}
\caption{Average slope of the age-metallicity degeneracy in different bins of stellar mass,
for the low-concentration and high-concentration galaxies of Figs~\ref{fig11} and 
\ref{fig12}.}\label{slopes}
\centering
\begin{tabular}{|l|cc|}
\multicolumn{3}{|c|}{}\\
\noalign{\smallskip}
\hline
Stellar mass ($M_\odot$) & \multicolumn{2}{|c|}{$\Delta \log Z/\Delta \log t_r$}\\
\noalign{\smallskip}
 & $C\leq2.4$ & $C\geq2.8$ \\
\noalign{\smallskip}
\hline
\noalign{\smallskip}
$\log M_\ast\leq10$	     & $-$1.78  & $-$1.29 \\
$10<\log M_\ast\leq10.3$    & $-$1.41  & $-$0.85 \\
$10.3<\log M_\ast\leq10.5$  & $-$1.28  & $-$0.84 \\
$10.5<\log M_\ast\leq10.8$  & $-$1.15  & $-$0.80 \\
$10.8<\log M_\ast\leq11$    & $-$1.05  & $-$0.78 \\
$\log M_\ast>11$	     & $-$0.95  & $-$0.75 \\
\noalign{\smallskip}
\hline
\end{tabular}
\end{table}

The orientation of the age-metallicity degeneracy in Figs~\ref{fig11} and 
\ref{fig12} indicates that it cannot be responsible for the trends in age and 
metallicity as a function of stellar mass identified above. However, it can
account for the shape of the joint age-metallicity probability contours for 
early-type galaxies and the most massive late-type galaxies at high metallicities
($Z\ga Z_\odot$). 

\subsection{Aperture bias}\label{aperture}
\begin{figure}
\centerline{\includegraphics[width=8truecm]{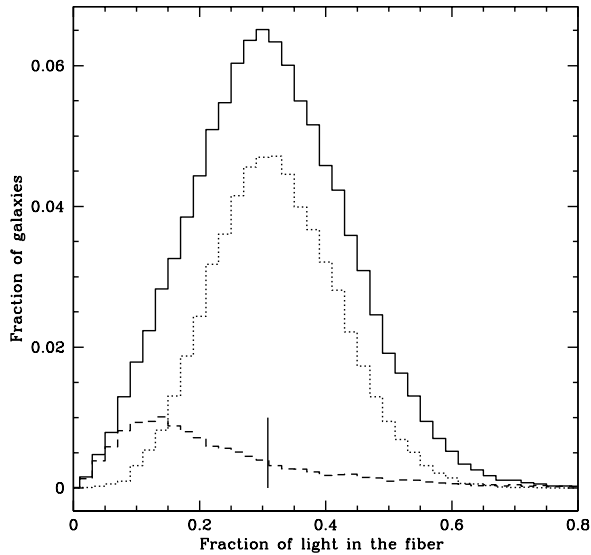}}
\caption{Distribution of the fraction of the total $r$-band flux collected by the SDSS
fibre, computed as the ratio between the fibre flux and the Petrosian flux, for 44,254
high-S/N galaxies in the SDSS DR2. The vertical bar indicates the median fraction for
this sample. The dashed and dotted lines show the distributions of low-concentration,
late-type galaxies and high-concentration, early-type galaxies, respectively.}
\label{fig13}
\end{figure}

The SDSS galaxy spectra on which our analysis is based are taken using 3\arcsec-diameter
fibres, which collect only the light coming from limited inner regions of the galaxies. 
We can estimate the fraction of the total galaxy light entering a fibre from the ratio
between the fibre flux and the total Petrosian flux. Fig.~\ref{fig13} shows the
distribution of this ratio for the 44,254 high-S/N galaxies in our sample. The median
value is about 30 percent, and the maximum value about 60 percent. The dashed and dotted
lines show the distributions for low-concentration, late-type galaxies and 
high-concentration, early-type galaxies, respectively. As expected, the fibre tends to
collect a smaller fraction of the total light for low-concentration galaxies than for 
high-concentration galaxies.

Radial gradients in the properties of stellar populations are known to exist for both
bulge-dominated \citep{henry99,saglia00,mehlert03} and disk-dominated \citep{BdJ00,
macarthur04} galaxies. Thus, aperture effects are a significant concern for our age and
metallicity estimates. Fortunately, the distribution of the fraction of total galaxy 
light that enters an SDSS fibre does not vary appreciably with redshift (except at very
low redshift), because larger galaxies are also brighter and hence selected out to larger
distances. 

One way to test for the effects of aperture bias on our results is to compare the ages
and metallicities of galaxies of similar luminosity located at different redshifts. 
Because of the presence of both a faint and a bright magnitude cut-off in the SDSS survey,
this comparison is possible only for galaxies in a limited absolute-luminosity range,
if we want to span a reasonably large redshift range. We consider galaxies with 
$r$-band absolute magnitudes in the narrow range $-21\leq r\leq-22$, which are observed at
redshifts between 0.02 and 0.12. At fixed absolute magnitude within this range, both low-
and high-concentration galaxies show a decrease (by $\sim0.15$ dex) in typical 
metallicity from redshift 0.02 to 0.12. High-concentration galaxies do not exhibit any
significant trend in age, but the typical age of low-concentration galaxies decreases 
by about 0.2 dex over the same redshift range.
  
Another way to examine the potential effects of aperture bias on our results is to 
compare the age and metallicity distributions of galaxies with similar properties as a
function of $z/z_{\rm max}$, where $z_{\rm max}$ is the minimum of the redshift at which
the galaxy reaches the survey faint-magnitude limit and that at which its median S/N per
pixel drops below 20 (we assume Poissonian noise, i.e. that the median S/N per pixel
scales as the square root of the apparent fibre luminosity). With this approach, we can
avoid restricting our test to galaxies in a narrow magnitude range. Also, since we are 
interested in possible effects of aperture bias on trends with galaxy mass, we divide 
our sample into the same 6 bins of stellar mass as in Figs~\ref{fig11} and 
\ref{fig12} above. 

\begin{figure*}
\centerline{\includegraphics[width=8truecm]{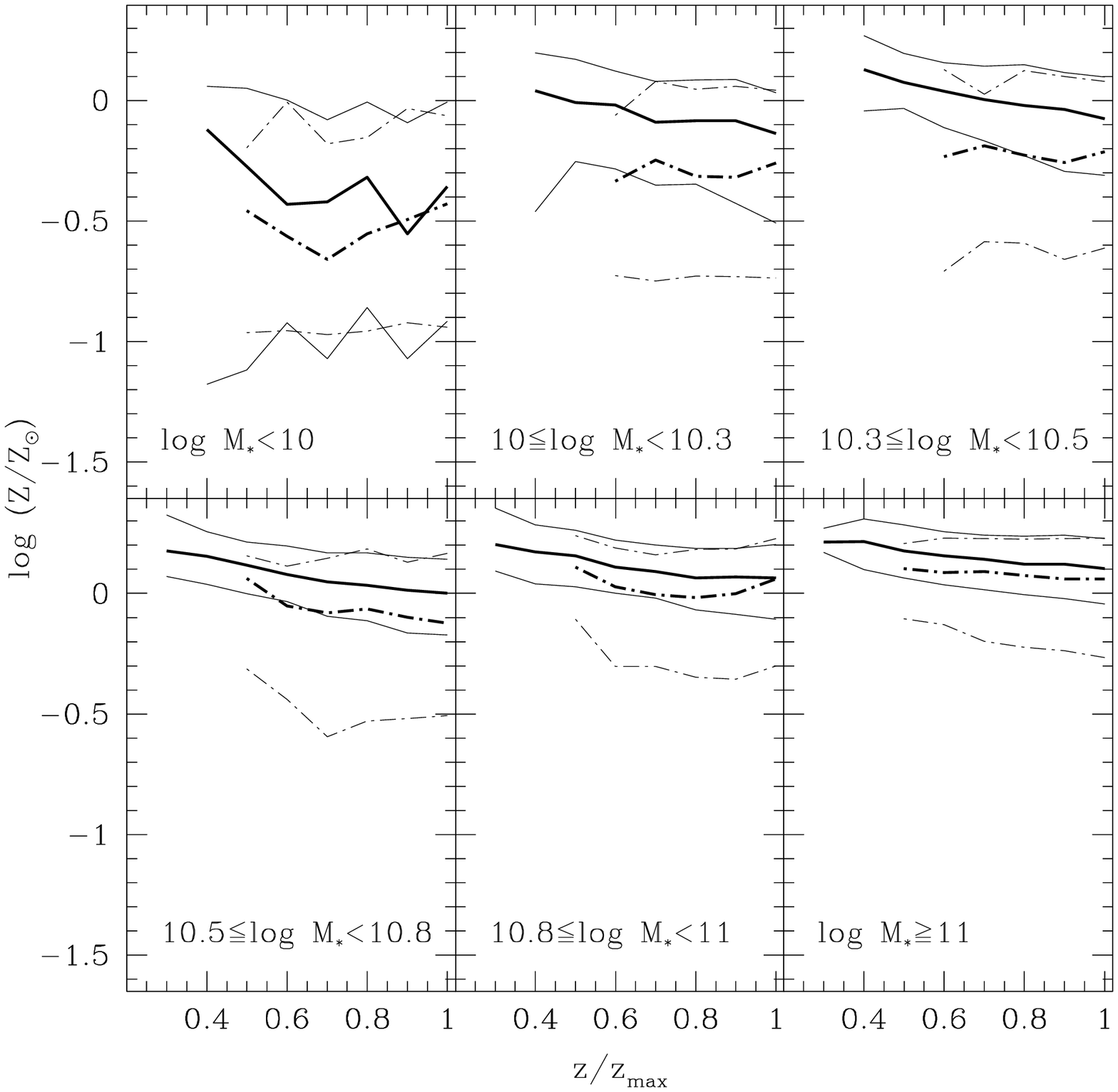}
\includegraphics[width=8truecm]{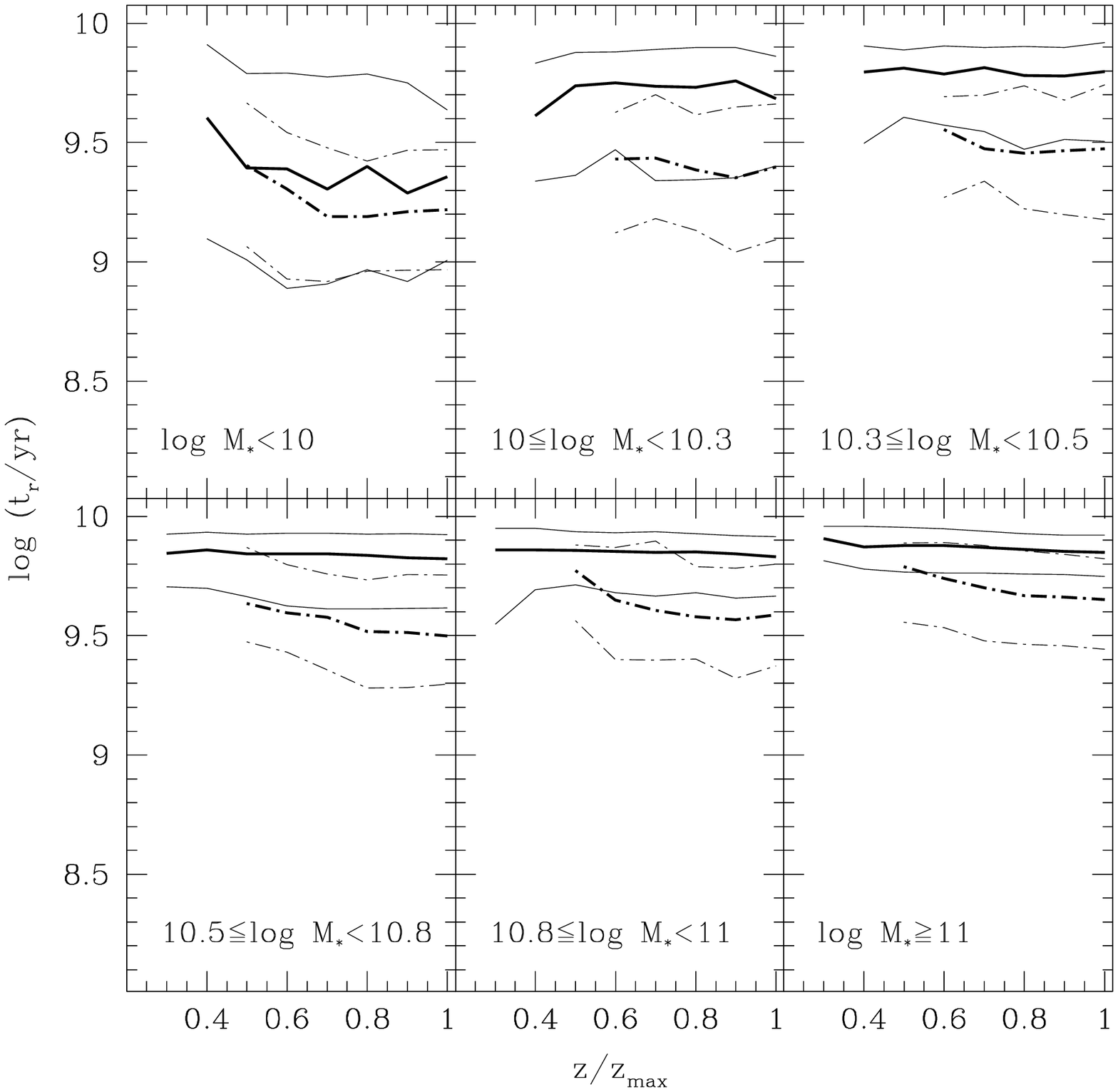}}
\caption{Dependence on $z/z_{\rm max}$ of the median (thick line) and the $16^{\rm th}$
and $84^{\rm th}$ percentiles (thin lines) of the distributions in stellar metallicity 
(left-hand plot) and age (right-hand plot) of SDSS-DR2 galaxies with median S/N per pixel
greater than 20. The quantity $z_{\rm max}$ is the minimum of the redshift at which the
galaxy reaches the survey magnitude limit and that at which its median S/N per pixel drops
below 20. The solid lines correspond to the subsample of 26,003 early-type ($C\geq2.8$) 
galaxies, while the dot-dashed lines correspond to the subsample of 5616 late-type 
($C\leq2.4$) galaxies.}
\label{fig14}
\end{figure*}

Fig.~\ref{fig14} shows the dependence on $z/z_{\rm max}$ of the median (thick line) and
the $16^{\rm th}$ and $84^{\rm th}$ percentiles (thin lines) of the distributions in
stellar metallicity (left-hand plot) and age (right-hand plot) of the high-S/N galaxies
in our sample. The solid lines correspond to the subsample of 26,003 early-type galaxies,
while the dot-dashed lines correspond to the subsample of 5616 late-type galaxies. For 
early-type galaxies of different masses, the largest change in median metallicity from 
one edge of the survey to the other is about 0.2 dex. This is comparable to the error 
associated with the metallicity estimates of most galaxies and smaller than the overall 
metallicity change with galaxy mass (see Fig.~\ref{fig8}). 
This variation is also consistent with the metallicity gradient of $-0.2$~dex per decade
in radius reported by several studies of bulge-dominated galaxies \citep[see][]{henry99,wu2004}.
We note that the largest variation arises for galaxies with intermediate stellar masses,
which often have both a prominent bulge and a disk. For disk-dominated, late-type 
galaxies, the trends in median metallicity with $z/z_{\rm max}$ are weaker than for 
early-type galaxies, although the distributions are broader.

The median age of early-type galaxies does not show any significant trend with 
$z/z_{\rm max}$ in any mass range in Fig.~\ref{fig14}. This is in agreement with 
earlier findings of negligible age gradients in such galaxies \citep{mehlert03,wu2004}. 
Late-type galaxies of low and intermediate mass show a similar behaviour. Only the
most massive late-type galaxies show a significant decrease (by about 0.2 dex) in
median age as a function of $z/z_{\rm max}$, as expected if age gradients in spiral
galaxies are stronger for larger galaxies \citep{macarthur04}.

The occurrence of negative radial age and metallicity gradients in different types of
galaxies and the fact that the SDSS spectra sample no more than 50--60 percent of the 
total light of a galaxy are likely to cause overestimates of the stellar metallicities
and ages of some galaxies in our sample. However, the weak trends in $\log Z$ and $\log
t_r$ as a function of $z/z_{max}$ in Fig.~\ref{fig14} suggest that aperture biases are
unlikely to have a significant effect on the trends we have found previously for 
metallicity and age as a function of stellar mass.
 
\section{Summary and conclusions}\label{summary}
We have used a new approach to derive estimates of light-weighted stellar metallicity, 
age and stellar mass from the optical spectra of a sample of $\sim$ 200,000 nearby 
galaxies ($\rm 0.005<z\leq0.22$) drawn from the SDSS DR2. Our method relies on the 
comparison of the galaxy spectra to a large library of model spectra at medium-high 
resolution, based on Monte Carlo star formation histories spanning the full physically
plausible range. Extending earlier work by \cite{kauf03a}, we have adopted a Bayesian
approach to derive the a posteriori likelihood distribution of each physical parameter
by computing the goodness of fit of the observed spectrum for all the models in the 
library. In practice, we compute only the strengths of a set of carefully selected spectral 
absorption features. 

Our analysis shows that stellar absorption features with different sensitivities to age
and metallicity must be fitted simultaneously in galaxy spectra to obtain good constraints
on both age and metallicity. We focus on several newly calibrated Lick indices and the
4000~{\AA} break (see below). The resolution of our models is higher than that used in
most previous studies and is well matched to the resolution of the SDSS spectra. This
allows us to measure the indices in the same way in model and galaxy spectra, avoiding 
any loss of information from the observed spectra. We emphasize that, while most previous
studies were restricted to the analysis of old stellar populations, our sample includes
galaxies in the full range of star formation activities. For actively star-forming 
galaxies, the higher resolution of the models is crucial to separate strong emission 
lines from the underlying stellar absorption.

Using this approach, we are able to constrain stellar metallicity and light-weighted age
to within $\pm0.15$ dex for the majority of the galaxies in the sample. Our stellar-mass 
estimates, determined within less than $\pm0.1$ dex for the whole sample, are in good 
agreement with those derived by \cite{kauf03a}. While the uncertainties on the stellar mass 
are almost independent of galaxy type, we find that, as expected, the smallest errors on 
age and especially metallicity are obtained for galaxies with the strongest absorption
lines.

The estimates we derive of galaxy parameters are affected by limitations of both model
and observed spectra. For example, the models do not include the effects of variations
in heavy-element abundance ratios, while systematic deviations from the solar 
abundance ratios are known to arise in external galaxies \citep[e.g.][]{wfg92}. When 
estimating metallicity, we therefore consider only those spectral features that have
been shown by previous studies to depend negligibly on $\alpha$/Fe abundance ratio
\citep{trager00b,Thomas03a, tantalo04}. The inclusion of higher-order Balmer lines, 
which may depend on element abundance ratio at high metallicities \citep{thomas04}, 
could lead to an overestimate of the metallicity and an underestimate of the age by 
$\sim$0.05 dex for galaxies with non-solar abundance ratios in our sample. Another systematic 
uncertainty in our metallicity and age estimates comes from the discrepancy between integrated 
colours and spectral index strengths for supra-solar metallicity models (the model spectral 
indices correspond to lower metallicities than the colours). Therefore, we expect metallicities 
inferred using these models to be biased toward high values at high metallicities. This problem
mainly affects massive, early-type galaxies with strong absorption lines. 

We have shown that, in addition to these model uncertainties, the signal-to-noise ratio
in a galaxy spectrum strongly affects our estimates of age and especially metallicity. 
We choose to rely primarily on results for galaxies with median S/N per pixel greater 
than 20. This cut reduces the original sample by about 75 percent and biases our analysis
toward high surface brightness, high concentration galaxies. Our main results appear 
unaffected by this cut in S/N; including lower-S/N galaxies mainly increases the 
scatter in the relations we find, because of the larger associated uncertainties on 
metallicity.

Another observational limitation is the small aperture sampled by the SDSS spectra.
The fraction of light collected by an SDSS fibre depends on the apparent size of the
target galaxy, and this fraction is less than 30 percent for most galaxies in our sample.
The stellar mass we derive from the spectrum has been scaled to the total luminosity of
the galaxy, and hence, it represents the total stellar mass of that galaxy. However, the
age and the metallicity derived from the fibre spectrum cannot be easily extrapolated
to total values without information about radial gradients. We have shown that these
two parameters vary weakly as a function of the fraction of light entering the 
fibre. This reassures us that aperture bias does not have a major effect on the 
relations between metallicity, age and stellar mass that we derive.   

We have explored how galaxies are distributed in metallicity, age and stellar mass in 
the diagram defined by 4000-\AA\ break and Balmer-line strength. As shown by 
\citet{kauf03a} the principal sequence in this plane reflects an increase in present 
to past-averaged star formation rate with decreasing \dn\ \citep[see also][]{jarle03}.
Galaxies with particularly strong Balmer absorption at given \dn\ have undergone recent 
starbursts. By including metallicity dependence in the modelling, we have shown that:

\begin{itemize}
\item
In this diagram, the average stellar mass increases along the sequence populated by
galaxies in our sample, from less than $10^{10} M_\odot$ to more than $10^{11}M_\odot$
from end to end of the sequence. The transition at intermediate \dn\ is smooth between
the two regimes. These results are in agreement with those of \cite{kauf03b}.
\item
Both stellar metallicity and light-weighted age also increase with \dn. Galaxies with
particularly strong Balmer absorption at intermediate 4000-\AA\ breaks ($\dn\sim1.5$)
are on average younger and more metal-rich than the bulk of the sequence. This supports
the idea that these galaxies may have recently undergone a burst of metal-enriched star
formation.
\end{itemize}

In addition, we have constructed the conditional distributions of metallicity and age 
estimates as a function of stellar mass. This analysis shows that:
\begin{itemize}
\item
Both stellar metallicity and light-weighted age increase with stellar mass, the
increase being rapid at intermediate masses. At masses above $\sim3\times10^{10}
M_\odot$, a gradual flattening occurs in both relations. This stellar mass 
corresponds to the transition mass identified by \cite{kauf03b} in plots of \dn,
surface mass density and concentration against stellar mass.
\item
Despite the above clear relations, metallicity and age are not uniquely determined by 
mass; there is an intrinsic scatter in both relations, which is largest at stellar masses
around $10^{10}M_\odot$. This scatter persists even when considering low- and 
high-concentration galaxies separately.
\item
The relation we find between stellar metallicity and stellar mass is similar to that
found by \cite{christy04} between gas-phase oxygen abundance and stellar mass. We 
confirm that higher stellar metallicities are indeed associated with higher gas-phase
metallicities. However, the relation between stellar and gas-phase metallicities has a
substantial scatter, which can only partly be attributed to the uncertainties in 
stellar-metallicity estimates. This suggests that a simple `closed-box' scenario is not
sufficient to explain the relation, and that a variety of gas accretion/ejection histories
may be required.
\end{itemize}

We have also explored relationships between metallicity, age and stellar mass for 
separate subsamples of high-concentration, early-type and low-concentration,
late-type galaxies. The distribution in age-metallicity space in bins of stellar 
mass shows that (Figs~\ref{fig11} and \ref{fig12}):
\begin{itemize}
\item
Relations between stellar metallicity, age and stellar mass hold for both early- and 
late-type galaxies. However, the age range of the sample is less mass-dependent for
early-type than for late-type galaxies.
\item
At masses below $\rm 10^{10} M_\odot$, late-type galaxies are younger and more metal-poor
than early-type galaxies. At masses above $\rm 10^{11} M_\odot$ the metallicities and 
ages of late-type galaxies are similar to those of early-type galaxies.
\end{itemize}

The above results indicate that young, metal-poor stellar populations are found 
predominantly in low-mass galaxies. This is consistent with a `downsizing' 
scenario, in which the mass and luminosity of the galaxies undergoing active star 
formation become progressively lower as the Universe becomes older \citep{cowie96}. 
Taken at face value, this would imply that the low metallicities of low-mass galaxies
just reflect the fact that these galaxies have had less time to produce
metals. This naive conclusion is not consistent with that of \citet{christy04}, who
showed that the effective yields (from indirect gas-mass fraction measurements) of
low-mass, star-forming galaxies in the SDSS DR2 are too low for these galaxies to be 
interpreted as young closed-box systems. This led \citet{christy04} to favor galatic
winds as the most likely origin of the relation between gas-phase metallicity and 
stellar mass. This interpretation is supported here by the fact that galaxies
with no \ha\ emission in our sample (which have presumably completed their star
formation) also show a trend of decreasing stellar metallicity with decreasing stellar
mass for $M_\ast\la 2\times10^{10}M_\odot$ (the metallicity of closed-box galaxies 
should tend to the yield near gas exhaustion, independently of mass). The large scatter
present in all relations between stellar metallicity, age 
and stellar mass (even when separated into late-type and early-type galaxies) 
further indicates that stellar mass is not the unique parameter determining the star 
formation history and hence the physical parameters of present-day galaxies. 
Additional variations in star formation and enrichment history (e.g. driven by 
gas infall or outflow) are required to explain these parameters.

An advantage of the large samples available from the SDSS is that they allow us to study
not only the mean relations between physical parameters of galaxies, but also the 
dependence of these parameters on various galaxy properties. Our results are particularly
useful for studying the physical origin of well-known observed relations, such as for
example the colour-magnitude relation and the relation between $\rm Mg_2$-index strength
and velocity dispersion for early-type galaxies. Our analysis also enables us to estimate
the total metal content of the local Universe and the distribution of these metals over
different galaxy types. We will address these questions in forthcoming papers.

A complete census of the physical parameters of galaxies out to $z\sim0.1$ is essential
for constraining models of the star formation and chemical enrichment histories of 
galaxies. The application of our method to large samples of galaxies at higher redshifts
will enable us to study how the distribution of the physical parameters of galaxies 
evolves with lookback time. Ongoing deep redshift surveys, such as VVDS \citep{lefevre04}
and DEEP2 \citep{deep2} in the optical and {\it GALEX} in the ultraviolet \citep{galex},
are already assembling large samples of high-redshift galaxies. Quantitative comparison
with the SDSS samples should substantially deepen our understanding of galaxy evolution. 
 
\section*{Acknowledgments}
We thank Guinevere Kauffmann and an anonymous referee for useful comments. A.G. thanks 
Stefano Zibetti and Gabriella De Lucia for useful discussions. A.G. and S.C. thank the 
Alexander von Humboldt Foundation, the Federal Ministry of Education and Research, and 
the Programme for Investment in the Future (ZIP) of the German Government for
funding through a Sofja Kovalevskaja award.  J.B. acknowledges the support of an 
ESA post-doctoral fellowship.

Funding for the creation and distribution of the SDSS Archive has been provided
by the Alfred P. Sloan Foundation, the Participating Institutions, the National
Aeronautics and Space Administration, the National Science Foundation, the US
Department of Energy, the Japanese Monbukagakusho, and the Max Planck Society.
The SDSS Web site is http://www.sdss.org/. The Participating Institutions are
the University of Chicago, Fermilab, the Institute for Advanced Study, the
Japan Participation Group, the Johns Hopkins University, the Max Planck
Institute for Astronomy (MPIA), the Max Planck Institute for Astrophysics
(MPA), New Mexico State University, Princeton University, the United States
Naval Observatory, and the University of Washington.

\label{lastpage}
\end{document}